\documentclass[preprint,amsfonts,amssymb,amsmath,eqsecnum,%
tightenlines,nofootinbib,showpacs,showkeys]{revtex4}

\newcommand{\cN}{\mathcal{N}}

\newcommand{\cV}{\mathcal{V}}
\newcommand{\cW}{\mathcal{W}}
\newcommand{\bA}{\boldsymbol{A}}
\newcommand{\bB}{\boldsymbol{B}}
\newcommand{\bC}{\boldsymbol{C}}
\newcommand{\bS}{\mathbf{S}}
\newcommand{\tH}{\tilde{H}}
\newcommand{\ee}{\mathrm{e}}
\newcommand{\dd}{\mathrm{d}}
\newcommand{\del}{\partial}

\newcommand{\rnu}{\sqrt{\nu}}
\newcommand{\fsl}{\mathfrak{sl}} 
 
\newcommand{\bbR}{\mathbb{R}}
\newcommand{\bbC}{\mathbb{C}}
\newcommand{\bbZ}{\mathbb{Z}}
\newcommand{\gd}{\operatorname{gd}}
\newcommand{\sn}{\operatorname{sn}}
\newcommand{\cn}{\operatorname{cn}}
\newcommand{\dn}{\operatorname{dn}}

\begin{document}

\title{$\fsl(M+1)$ Construction of Quasi-solvable Quantum
$M$-body Systems}
\author{Toshiaki Tanaka}
\email{ttanaka@mail.tku.edu.tw}
\affiliation{Department of Physics,
Graduate School of Science,\\
Osaka University, Toyonaka,
Osaka 560-0043, Japan}
\altaffiliation{Present address: Department of Physics,
Tamkang University, Tamsui 25137, Taiwan, R.O.C.}


\begin{abstract}
We propose a systematic method to construct quasi-solvable quantum
many-body systems having permutation symmetry. By the introduction of
elementary symmetric polynomials and suitable choice of a
solvable sector, the algebraic structure of $\fsl(M+1)$ naturally
emerges. The procedure to solve the canonical-form condition for the
two-body problem is presented in detail. It is shown that the resulting
two-body quasi-solvable model can be uniquely generalized to the
$M$-body system for arbitrary $M$ under the consideration of the
$GL(2,K)$ symmetry.
An intimate relation between quantum solvability and supersymmetry
is found. With the aid of the $GL(2,K)$ symmetry, we classify the
obtained quasi-solvable quantum many-body systems. It turns out that
there are essentially five inequivalent models of Inozemtsev type.
Furthermore, we discuss the possibility of including $M$-body
($M\ge 3$) interaction terms without destroying the quasi-solvability.
\end{abstract}

\pacs{02.30.Ik; 03.65.Ca; 03.65.Fd; 03.65.Ge}
\keywords{quantum many-body problem; quasi-solvability; supersymmetry;
Inozemtsev models; Calogero--Sutherland models}

\preprint{OU-HEP 435}

\maketitle

\section{\label{sec:intro}Introduction}

Since the discovery of quasi-solvability in one-dimensional quantum
mechanics~\cite{TuUs1}, many new ideas and concepts have been
discovered and developed. The progress until the early 1990s is
summarized in Ref.~\cite{Ushve}.
So far, most of the known quasi-solvable
quantum systems of one-degree of freedom are related to the ones
constructed from the $\fsl(2)$ generators~\cite{Turbi1}. This family
of quasi-solvable models was completely classified under the
consideration of the $GL(2,\bbR)$ invariance of the
models~\cite{LoKaOl3,LoKaOl4}. Recently, a new type of orthogonal
polynomial systems was discovered in connection with a quasi-solvable
model~\cite{BeDu1}. Soon later, it was shown that similar type of
polynomial systems, known as weakly orthogonal polynomial systems,
can be constructed for all the $\fsl(2)$ quasi-exactly solvable
models~\cite{KrUsWa1,FiLoRo1}. In the last couple of years, much attention
has been paid to quasi-solvable models from the viewpoint of $\cN$-fold
supersymmetry~\cite{AKOSW2,ASTY,AST1,KlPl2,KlPl3,DoDuTa1,DoDuTa2,AST2,%
ANST1,ANST2,ST1,CaIoNi1,KlPl5,Tanak3}.
Furthermore, several realistic physical systems have been found,
which can be reduced to one-dimensional quasi-solvable
models~\cite{KlPl3,SaGh1,Taut1,Taut2,Turbi3,Taut3,ViPi1,Taut4,HoKh1,%
ChHo1,ChHo2}.

Unlike one-body systems, construction of quasi-solvable
quantum many-body systems suffers from another difficult problem.
The difficulty originates from the fact that second-order
differential operators of several variables are, in general,
no longer equivalent to the Schr\"{o}dinger operators. Indeed,
construction of quasi-solvable two-body problems by the naive extension
of the $\fsl(2)$ method to the rank two algebras, investigated
extensively in the early 1990s~\cite{Ushve,LoKaOl4,ShTu1,Shifm1,%
LoKaOl1,LoKaOl2,LoKaOl5,Turbi2}, led to the Schr\"{o}dinger operators
in curved space.
In 1995, a breakthrough was achieved in Ref.~\cite{RuTu1}, where the
exact solvability of the rational and trigonometric $A$ type
Calogero--Sutherland (CS) models~\cite{Calog1,Suthe1,Moser1} for any
finite number of particles were shown by a similar algebraic method.
The key ingredient is to construct the generators of $\fsl(M+1)$
Lie algebra in terms of the elementary symmetric polynomials which
reflect the permutation symmetry of the CS models. This idea was
further employed to show the exact solvability of the rational and
trigonometric $A$ and $BC$ type CS models and their supersymmetric
generalizations~\cite{BrTuWy1}, and to show the quasi-exact solvability
of various deformed CS models~\cite{MiRoTu1,HoSh1}. So, it is natural
to ask what kinds of quasi-solvable quantum many-body systems can
be constructed from the same algebraic procedure. Recently, this
classification problem was partly accessed in Ref.~\cite{UlLoRo1}
though the method depends on the specific ansatz employed.
In our previous short letter~\cite{Tanak1}, we have briefly reported
a partial answer for this problem without recourse to any
specific ansatz by restricting the consideration up to two-body
interactions. In this paper, we will give the details and the
further developments on the issues. Especially, we will see that the
$GL(2,K)$ symmetry plays a central role not only on the classification
problem but also on the problem of what kind of many-body operators
can exist in the quasi-solvable quantum Hamiltonian.

The paper is organized as follows. In the next section, we give the
definition of quasi-solvability together with its important subclasses
such as quasi-exact solvability, solvability and so on. Based on
the definition, we present a procedure to construct a quasi-solvable
model by introducing a set of differential operators. In
Section~\ref{sec:algeb}, we show that, by the introduction of
elementary symmetric polynomials and suitable choice of the set
of differential operators which determines a solvable sector of the
model, a family of quasi-solvable models can be constructed from
a polynomial of the $\fsl(M+1)$ generators of second-degree.
Section~\ref{sec:canon} deals with the canonical-form condition,
namely, under what conditions we can obtain the quasi-solvable
\textit{quantum systems} from the algebraic operators constructed
in Section~\ref{sec:algeb}. We explicitly show the procedure to
solve the canonical-form condition for the two-body problem.
In Section~\ref{sec:gl2ks}, we explain the $GL(2,K)$ symmetry of
the model. It turns out that this symmetry is so restrictive that
the quasi-solvable quantum two-body model obtained in
Section~\ref{sec:canon} can be uniquely generalized to $M$-body ($M\ge 3$)
systems. An interesting relation between (quasi-)solvability and
supersymmetry is discussed. In Section~\ref{sec:inver}, we show that
the many-body quantum systems obtained solely by the consideration
of the $GL(2,K)$ symmetry are actually quasi-solvable. This is achieved
by expressing the models in terms of the elementary symmetric polynomials.
In Section~\ref{sec:class}, we classify the models under the
consideration of the $GL(2,K)$ symmetry, by essentially the same way
as in Ref.~\cite{LoKaOl3}. The explicit forms of both the superpotentials
and potentials are presented. In Section~\ref{sec:possi}, we investigate
the possibility of including $M$-body ($M\ge 3$) interaction terms without
destroying the quasi-solvability. Again, the $GL(2,K)$ symmetry plays
an essential role in this problem. We give several concluding
remarks in the final section. Some useful formulae are summarized in
Appendices.

\section{\label{sec:qsqmp}Quasi-solvability in Many-body Systems}

First of all, we will give the definition of quasi-solvability and
some notions of its special cases based on Refs.~\cite{Turbi2,AST2}.
A linear differential operator $H$ of several variables
$q=(q_{1},\dots,q_{M})$ is said to be \textit{quasi-solvable}
if it preserves a finite dimensional functional space $\cV_{\cN}$
whose basis admits an explicit analytic form:
\begin{align}
H\cV_{\cN}&\subset\cV_{\cN}, & \dim\cV_{\cN}&=n(\cN)<\infty,
 & \cV_{\cN}&=\textrm{span}\,\left\{\phi_{1}(q),\dots,\phi_{n(\cN)}(q)
 \right\}.
\label{eqn:defqs}
\end{align}
An immediate consequence of the above definition of quasi-solvability
is that, since we can calculate finite dimensional matrix elements
$\bS_{k,l}$ defined by,
\begin{align}
H\phi_{k}=\sum_{l=1}^{n(\cN)}\bS_{k,l}\phi_{l}\qquad
 \bigl(k=1,\dots,n(\cN)\bigr),
\label{eqn:defbS}
\end{align}
we can diagonalize the operator $H$ and obtain its spectra
on the space $\cV_{\cN}$, at least, algebraically. Furthermore,
if the space $\cV_{\cN}$ is a subspace of a Hilbert space
$L^{2}(S)$ $(S\subset\bbR^{M})$ on which the operator $H$ is naturally
defined, the solvable spectra and the corresponding vectors of
$\cV_{\cN}$ give the \textit{exact} eigenvalues and eigenfunctions of
$H$, respectively. In this case, the operator $H$ is said to be
\textit{quasi-exactly solvable} (on $S$). Otherwise, the solvable
spectra and the corresponding vectors of $\cV_{\cN}$ only give
\textit{local} solutions of the characteristic equation and have, at
most, restrictive meanings in the perturbation theory defined on the
physical space $L^{2}(S)$~\cite{AKOSW2,AST2,ST1}.

A quasi-solvable operator $H$ of several variables
is said to be \textit{solvable}
if it preserves an infinite flag of finite dimensional functional
spaces $\cV_{\cN}$,
\begin{align}
\cV_{1}\subset\cV_{2}\subset\dots\subset\cV_{\cN}\subset\cdots,
\label{eqn:flagv}
\end{align}
whose bases admit explicit analytic forms, that is,
\begin{align}
H\cV_{\cN}&\subset\cV_{\cN}, & \dim\cV_{\cN}&=n(\cN)<\infty,
 & \cV_{\cN}&=\textrm{span}\,\left\{\phi_{1}(q),\dots,\phi_{n(\cN)}(q)
 \right\},
\label{eqn:defsv}
\end{align}
for $\cN=1,2,3,\ldots$.
Furthermore, if the sequence of the spaces (\ref{eqn:flagv}) defined
on $S\subset\bbR^{M}$ satisfies,
\begin{align}
\overline{\cV_{\cN}(S)}\rightarrow L^{2}(S)\qquad
 (\cN\rightarrow\infty),
\label{eqn:limvs}
\end{align}
the operator $H$ is said to be \textit{exactly solvable} (on $S$).

In the case of a system of a single variable, it was
proved that quasi-solvability is essentially equivalent to
\textit{$\cN$-fold supersymmetry}~\cite{AST2}. In
this context, it is convenient to formulate the quasi-solvability
in terms of $\cN$th-order linear differential operators. Let us
introduce $\cN$-fold supercharges by,
\begin{align}
Q_{\cN}=\sum_{\{i\}}P_{\cN}^{\{i\}\dagger}\psi_{\{i\}},
 \qquad Q_{\cN}^{\dagger}=\sum_{\{i\}}P_{\cN}^{\{i\}}
 \psi_{\{i\}}^{\dagger},
\label{eqn:defsc}
\end{align}
where $\psi_{\{i\}}$ and $\psi_{\{i\}}^{\dagger}$ are
fermionic coordinates and $P_{\cN}^{\{i\}}$ are
$\cN$th-order linear differential operators of several variables
$q_{i}$. Using these operators, we define the vector space $\cV_{\cN}$
as,
\begin{align}
\cV_{\cN}=\bigcap_{\{i\}}\ker P_{\cN}^{\{i\}}.
\label{eqn:cvker}
\end{align}
Now, it is easy to see that an operator $H$ is quasi-solvable
with the solvable sector (\ref{eqn:cvker}) if the following
quasi-solvability condition holds~\cite{AST2}:
\begin{align}
P_{\cN}^{\{i\}}H\cV_{\cN}=0
 \quad\forall\,\{i\}.
\label{eqn:anoqs}
\end{align}
Therefore, the above formulation gives a concrete procedure to
construct a quasi-solvable operator.
In practice, however, the problem often becomes more tractable when
we make a suitable similarity transformation on the vector space
and on the operators and make a suitable change of variables.
In the case of $M=1$, the procedure was successfully employed
to construct the type A $\cN$-fold supersymmetric models
\cite{ANST1,ANST2,Tanak3}.
In the following section, we will generalize the procedure to the
case of several variables.

\section{\label{sec:algeb}$\fsl(M+1)$ Algebraization of the Models}

Let us consider an $M$-body quantum Hamiltonian for a system of
identical particles on a line,
\begin{align}
H_{\cN}=-\frac{1}{2}\sum_{i=1}^{M}\frac{\partial^{2}}%
{\partial q_{i}^{2}}+V(q_{1},\dots,q_{M}),
\label{eqn:psham}
\end{align}
which possesses permutation symmetry, that is,
\begin{align}
V({}\dots,q_{i},\dots,q_{j},\dots)=
V({}\dots,q_{j},\dots,q_{i},\dots)
\quad\forall\, i\neq j.
\label{eqn:pspot}
\end{align}
To algebraize the Hamiltonian (\ref{eqn:psham}), we will proceed
the following three steps. First, we make a \textit{gauge}
transformation on the Hamiltonian (\ref{eqn:psham}):
\begin{align}
\tH_{\cN}=\ee^{\cW(q)}H_{\cN}\,\ee^{-\cW(q)}.
\label{eqn:gtham}
\end{align}
The function $\cW(q)$ is to be determined later and plays the role
of the superpotential when the system Eq.~(\ref{eqn:psham}) is
supersymmetric. As in Eq.~(\ref{eqn:gtham}), we will hereafter
attach tildes to both operators and vector spaces to indicate
that they are quantities gauge-transformed from the original ones.
In the next step, we change the variables $q_{i}$ to $h_{i}$ by
a function $h$ of a single variable; $h_{i}=h(q_{i})$.
Note that the way of changing the variables preserves
the permutation symmetry. The third step is the introduction of
elementary symmetric polynomials of $h_{i}$ defined by,
\begin{align}
\sigma_{k}(h)=\sum_{i_{1}<\dots <i_{k}}^{M}h_{i_{1}}\cdots h_{i_{k}}
\quad (k=1,\dots,M),
\label{eqn:sigma}
\end{align}
from which we further change the variables to $\sigma_{i}$.
Owing to the permutation symmetry, the gauged Hamiltonian
(\ref{eqn:gtham}) can be completely expressed in terms of these
elementary symmetric polynomials (\ref{eqn:sigma}).
We choose a set of components of the $\cN$-fold supercharges in terms
of the above variables $\sigma_{i}$ as follows:
\begin{align}
\tilde{P}_{\cN}^{\{i\}}=\frac{\partial^{\cN}}{
 \partial\sigma_{i_{1}}\cdots\partial\sigma_{i_{\cN}}}
 \quad (1\le i_{1},\dots,i_{\cN}\le M),
\label{eqn:cptsc}
\end{align}
where $\{i\}$ is an abbreviation of the set $\{i_{1},\dots,i_{\cN}\}$.
Using these $\cN$-fold supercharges, we define the vector space
by Eq.~(\ref{eqn:cvker}), in which,
\begin{align}
\tilde{\cV}_{\cN}
 =\mathrm{span}\,\left\{\sigma_{1}^{n_{1}}\cdots\sigma_{M}^{n_{M}}:
 n_{i}\in\bbZ_{\ge 0},\, 0\le\sum_{i=1}^{M}n_{i}\le\cN -1\right\}.
\label{eqn:svspc}
\end{align}
For given $M$ and $\cN$, the dimension of the vector space
(\ref{eqn:svspc}) becomes,
\begin{align}
\dim \tilde{\cV}_{\cN}=\sum_{n=0}^{\cN -1}\frac{(n+M-1)!}{n!\, (M-1)!}
=\frac{(\cN +M-1)!}{(\cN -1)!\, M!}.
\label{eqn:dimvs}
\end{align}
We will construct the system (\ref{eqn:gtham}) to be quasi-solvable
so that the solvable subspace is given by just
Eq.~(\ref{eqn:svspc}). This can be achieved by imposing the
quasi-solvability condition of the gauge-transformed
form~\cite{AST2,ANST1},
\begin{align}
\tilde{P}_{\cN}^{\{i\}}\tH_{\cN}\tilde{\cV}_{\cN}=0
\quad \forall\,\{i\}.
\label{eqn:qscon}
\end{align}
The above condition ensures $\tH_{\cN}\tilde{\cV}_{\cN}
\subset\tilde{\cV}_{\cN}$, that is, the subspace $\tilde{\cV}_{\cN}$
is invariant under the action of $\tH_{\cN}$.
As a consequence, we can diagonalize the gauged Hamiltonian
(\ref{eqn:gtham}) in the subspace (\ref{eqn:svspc}) with the finite
dimension given by Eq.~(\ref{eqn:dimvs}), and can obtain the
corresponding spectra algebraically. Then, it is evident that
they are also the spectra of the original Hamiltonian
(\ref{eqn:psham}) in the subspace $\cV_{\cN}$ given by,
\begin{align}
\cV_{\cN}=\mathrm{span}\,\left\{\sigma_{1}^{n_{1}}\cdots
 \sigma_{M}^{n_{M}}\ee^{-\cW(q)}: n_{i}\in\bbZ_{\ge 0},\,
 0\le\sum_{i=1}^{M}n_{i}\le\cN -1\right\}.
\label{eqn:svosp}
\end{align}
Thus, the system $H_{\cN}$ turns to be quasi-solvable.

The general solution of Eq.~(\ref{eqn:qscon}) can be obtained in
the same way as in Ref.~\cite{ANST1}. As in the case of the one-body
models, it is sufficient to find differential operators up to
second-order as solutions for $\tH_{\cN}$ since we are
constructing a Schr\"{o}dinger operator in the original variables
$q_{i}$. We can find that solutions expressed in terms of first-order
derivatives are the following:
\begin{subequations}
\label{eqns:gensl}
\begin{align}
\frac{\partial}{\partial\sigma_{i}}&\equiv E_{0i},
\label{eqn:gensl0}\\
\sigma_{i}\frac{\partial}{\partial\sigma_{j}}&\equiv E_{ij},
\label{eqn:gensl1}\\
\sigma_{i}\left(\cN -1-\sum_{k=1}^{M}\sigma_{k}
 \frac{\partial}{\partial\sigma_{k}}\right)&\equiv
 \sigma_{i} E_{00}\equiv E_{i0}.
\label{eqn:gensl2}
\end{align}
\end{subequations}
These $M^{2}+2M$ operators $E_{\kappa\lambda}(\kappa+\lambda>0)$
together with $E_{00}$ constitute the $\mathfrak{gl}(M+1)$
Lie algebra:
\begin{align}
\left[ E_{\kappa\lambda}, E_{\mu\nu}\right]=
 \delta_{\mu,\lambda}E_{\kappa\nu}-\delta_{\kappa,\nu}E_{\mu\lambda}.
\label{eqn:slalg}
\end{align}
Here we note that throughout this paper the Latin indices take values
from $1$ to $M$ while Greek indices from $0$ to $M$, as in the standard
convention in relativistic theories.
Solutions involving second-order derivatives are of the following
form:
\begin{subequations}
\label{eqns:2ndsls}
\begin{align}
\frac{\del^{2}}{\del\sigma_{i}\del\sigma_{j}}&=E_{0i}E_{0j},\\
\sigma_{i}\frac{\del^{2}}{\del\sigma_{j}\del\sigma_{k}}&=E_{ij}E_{0k},\\
\sigma_{i}\sigma_{j}\frac{\del^{2}}{\del\sigma_{k}\del\sigma_{l}}
 &=E_{ik}E_{jl}-\delta_{j,k}E_{il},\\
\sigma_{i}\biggl(\cN-1-\sum_{l=1}^{M}\sigma_{l}\frac{\del}{\del\sigma_{l}}
 \biggr)\sigma_{j}\frac{\del}{\del\sigma_{k}}&=E_{i0}E_{jk},\\
\sigma_{i}\sigma_{j}\biggl(\cN-2-\sum_{k=1}^{M}\sigma_{k}
 \frac{\del}{\del\sigma_{k}}\biggr)\biggl(\cN-1-\sum_{l=1}^{M}\sigma_{l}
 \frac{\del}{\del\sigma_{l}}\biggr)&=E_{i0}E_{j0}.
\end{align}
\end{subequations}
Therefore, the general solution which contains up to the
second-order differential operators reads,
\begin{align}
\tH_{\cN}=-
\!\!\!
\sum_{\kappa,\lambda,\mu,\nu=0}^{M}
\!\!\!
A_{\kappa\lambda,\mu\nu}E_{\kappa\lambda}E_{\mu\nu}
+\sum_{\kappa,\lambda=0}^{M}B_{\kappa\lambda}E_{\kappa\lambda}
-C,
\label{eqn:gsham}
\end{align}
where $A_{\kappa\lambda,\mu\nu}$, $B_{\kappa\lambda}$, $C$ are
arbitrary constants. Since $E_{00}+\sum_{i=1}^{M}E_{ii}=\cN-1$,
the algebra represented by Eqs.~(\ref{eqns:gensl}) is essentially
$\fsl(M+1)$ and we can set $A_{\kappa\lambda,00}=A_{00,\mu\nu}=0$
$\forall\,\kappa,\lambda,\mu,\nu$ and $B_{00}=0$.
However, all the terms in Eq.~(\ref{eqn:gsham}) are not linearly
independent and thus we can set some of the constants zero.
From the Lie algebraic relations (\ref{eqn:slalg}), we can put,
\begin{subequations}
\label{eqns:redun1}
\begin{align}
A_{i0,j0}=A_{i0,jl}=0\quad&\forall\, i<j,\\
A_{ik,jl}=0\quad&\forall\, i<j\ \text{or}\ \forall\, k<l,\\
A_{ik,0l}=A_{0k,0l}=0\quad&\forall\, k<l,\\
A_{0k,jl}=A_{0k,j0}=A_{ik,j0}=0\quad&\forall\, i,j,k,l.
\label{eqn:redun14}
\end{align}
\end{subequations}
Furthermore, we can remove the remaining redundant terms
as follows:
\begin{align}
E_{i0}E_{0l}=-\sum_{k=1}^{M}E_{ik}E_{kl}+(M-\cN+1)E_{il}
 \quad\Rightarrow\quad
 A_{i0,0l}=0\quad\forall\, i,l.
\label{eqn:redun2}
\end{align}
After the substitutions (\ref{eqns:redun1}) and (\ref{eqn:redun2}),
the remaining terms in Eq.~(\ref{eqn:gsham}) turn to be linearly
independent.
Substituting the generators (\ref{eqns:gensl}) for
Eq.~(\ref{eqn:gsham}) and taking Eqs.~(\ref{eqns:redun1}) and
(\ref{eqn:redun2}) into account, we obtain the general solution
(\ref{eqn:gsham}) in terms of $\sigma_{i}$:
\begin{align}
\tH_{\cN}=&\,-\sum_{k,l=1}^{M}\bigl[\bA_{0}(\sigma)
 \sigma_{k}\sigma_{l}-\bA_{k}(\sigma)\sigma_{l}+\bA_{kl}(\sigma)\bigr]
 \frac{\partial^{2}}{\partial\sigma_{k}\partial\sigma_{l}}\notag\\
&\,+\sum_{k=1}^{M}\bigl[\bB_{0}(\sigma)\sigma_{k}-\bB_{k}(\sigma)
 \bigr]\frac{\partial}{\partial\sigma_{k}}-\bC(\sigma),
\label{eqn:sgham}
\end{align}
where $\bA_{\kappa}$, $\bA_{kl}$, $\bB_{\kappa}$ and $\bC$ are
the following second-degree polynomials of several variables:
\begin{subequations}
\label{eqns:sopols}
\begin{align}
\bA_{\kappa}(\sigma)&=\sum_{i\ge j}^{M}A_{i0,j\kappa}
 \sigma_{i}\sigma_{j},
\label{eqn:sopol1}\\
\bA_{kl}(\sigma)&=\sum_{i\ge j}^{M}A_{ik,jl}\sigma_{i}\sigma_{j}
 +\sum_{i=1}^{M}A_{ik,0l}\sigma_{i}+A_{0k,0l}\qquad (k\ge l),
\label{eqn:sopol2}\\
\bB_{0}(\sigma)&=2(\cN-2)\bA_{0}(\sigma)-\sum_{i=1}^{M}
 B_{i0}\sigma_{i},
\label{eqn:sopol3}\\
\bB_{k}(\sigma)&=(\cN-2)\bA_{k}(\sigma)+\sum_{i\ge j(\ge k)}^{M}
 A_{ij,jk}\sigma_{i}-\sum_{i=1}^{M}B_{ik}\sigma_{i}-B_{0k},
\label{eqn:sopol4}\\
\bC(\sigma)&=(\cN-1)(\cN-2)\bA_{0}(\sigma)-(\cN-1)\sum_{i=1}^{M}
 B_{i0}\sigma_{i}+C.
\label{eqn:sopol5}
\end{align}
\end{subequations}

Before closing the section, we will investigate the condition under
which the quasi-solvable operator (\ref{eqn:sgham}) becomes solvable.
If we act the $\fsl(M+1)$ generators for a fixed $\cN$ on the vector
spaces $\tilde{\cV}_{k}$ for any $k=1,2,3,\ldots$, we have,
\begin{align}
E_{0i}\tilde{\cV}_{k}\subset\tilde{\cV}_{k-1},\qquad
 E_{ij}\tilde{\cV}_{k}\subset\tilde{\cV}_{k},\qquad
 E_{i0}\tilde{\cV}_{k}\subset\tilde{\cV}_{k+1}.
\label{eqn:grade}
\end{align}
Thus, the operator (\ref{eqn:gsham}) preserves an infinite flag
of the spaces $\tilde{\cV}_{k}$ given by Eq.~(\ref{eqn:svspc}),
\begin{align}
\tilde{\cV}_{1}\subset\tilde{\cV}_{2}\subset\dots\subset
 \tilde{\cV}_{k}\subset\cdots,
\label{eqn:gifls}
\end{align}
if it does not contain the operators such as $E_{i0}E_{j0}$,
$E_{i0}E_{jk}$ and $E_{i0}$ at all. If we note from
Eqs.~(\ref{eqn:redun14}) and (\ref{eqn:redun2}) that we can arrange
the operator (\ref{eqn:gsham}) so that it does not contain operators
like $E_{0k}E_{j0}$ and $E_{i0}E_{0l}$, we conclude that the operator
is solvable if it does not contain the \textit{raising} operator
$E_{i0}$ at all. Thus, the solvability condition can be written as,
\begin{align}
A_{i0,j\kappa}=B_{i0}=0\quad\forall\, i,j,\kappa.
\label{eqn:solc1}
\end{align}
Finally, if we substitute the above condition for the set of polynomials
(\ref{eqns:sopols}), we get the solvability condition for the gauged
Hamiltonian (\ref{eqn:sgham}) as,
\begin{align}
\bA_{\kappa}(\sigma)=\bB_{0}(\sigma)=0 \quad\forall\,\kappa.
\label{eqn:solc21}
\end{align}
If this is the case, $\bB_{k}(\sigma)$'s become polynomials of at most
first-degree while $\bC(\sigma)$ becomes a constant:
\begin{align}
\bB_{k}(\sigma)&=\sum_{i\ge j(\ge k)}^{M}
 A_{ij,jk}\sigma_{i}-\sum_{i=1}^{M}B_{ik}\sigma_{i}-B_{0k},
\label{eqn:solc22}\\
\bC(\sigma)&=C.
\label{eqn:solc23}
\end{align}

\section{\label{sec:canon}Canonical-form Condition}

It is evident from the construction that any operators obtained from
Eq.~(\ref{eqn:sgham}) by gauge transformations are quasi-solvable.
However, this does not necessarily fulfil our purpose at the quantum
level. One of the most difficult problems one would encounter
in the algebraic approach to the quasi-solvable \textit{quantum}
many-body systems is to solve the canonical-form condition:
\begin{align}
H_{\cN}=\ee^{-\cW(q)}\tH_{\cN}\,\ee^{\cW(q)}=-\frac{1}{2}
 \sum_{i=1}^{M}\frac{\partial^{2}}{\partial q_{i}^{2}}+V(q).
\label{eqn:cfcon}
\end{align}
If the Hamiltonian (\ref{eqn:sgham}) is gauge-transformed
back to the original one, it in general does not take the canonical
form of the Schr\"odinger operator like Eq.~(\ref{eqn:psham})
and one can hardly solve, for arbitrary $M$, the conditions
under which a gauge-transform of Eq.~(\ref{eqn:sgham})
could be cast in the Schr\"odinger form. This difficulty can,
however, be overcome in our case by considering the symmetries.

Suppose we can solve the canonical-form condition for an $M$.
Then we would obtain a quasi-solvable Hamiltonian which might
contain up to $M$-body interactions:
\begin{align}
H_{\cN}=&\,-\frac{1}{2}\sum_{i=1}^{M}\frac{\partial^{2}}{
 \partial q_{i}^{2}}+g_{1}(M)\sum_{i=1}^{M}V_{1}(q_{i})
 +g_{2}(M)\sum_{i<j}^{M}V_{2}(q_{i},q_{j})+\cdots\notag\\
&\,\cdots+g_{m}(M)\sum_{i_{1}<\dots<i_{m}}^{M}
 V_{m}(q_{i_{1}},\dots,q_{i_{m}})+\dots+g_{M}(M)
 V_{M}(q_{1},\dots,q_{M}),
\label{eqn:mbqs1}
\end{align}
where each $g_{m}(M)(1\le m\le M)$ denotes the coupling constant of
the corresponding $m$-body interaction and may depend on the
number of the particle $M$. 
Then, we can get a quasi-solvable $M$-body model with up to
$M'$-body interactions $(M'<M)$ if we turn off all the coupling
constants $g_{m}(M)$ for $m>M'$:
\begin{align}
H_{\cN}=&\,-\frac{1}{2}\sum_{i=1}^{M}\frac{\partial^{2}}{
 \partial q_{i}^{2}}+g_{1}(M)\sum_{i=1}^{M}V_{1}(q_{i})
 +g_{2}(M)\sum_{i<j}^{M}V_{2}(q_{i},q_{j})+\cdots\notag\\
&\,\cdots+g_{M'}(M)\sum_{i_{1}<\dots<i_{M'}}^{M}
 V_{M'}(q_{i_{1}},\dots,q_{i_{M'}}).
\label{eqn:mbqs2}
\end{align}
The resultant model should be, when we put $M=M'$, one of the
models constructed from the $\fsl(M'+1)$. This is a consequence
of the permutationally symmetric construction. Actually, the
gauged Hamiltonian (\ref{eqn:gsham}) composed of the $\fsl(M+1)$
generators is reduced to the one composed of the generators
of the subalgebra $\fsl(M'+1)\subset\fsl(M+1)$ if we set
$h_{i}=0$ for $i>M'$ involved in both the generators (\ref{eqns:gensl})
and the representation space (\ref{eqn:svspc}).

The above observation means that, we can know the functional form
of the $M$-body quasi-solvable potential which involves up to
$M'$-body interactions $(M'<M)$ if we can solve the canonical-form
condition for the $M'$-body case. Furthermore, as we will later see
in section \ref{sec:gl2ks}, the $M$-dependence of each coupling
constant $g_{m}(M)$ can be uniquely determined by considering another
symmetry, namely, $GL(2,K)$ symmetry. Therefore, it is necessary and
sufficient to solve the $M'$-body canonical-form condition as far
as we are concerned with the quasi-solvable models containing up to
$M'$-body interactions. The simplest but non-trivial case is the
two-body case. In the followings, we will see that we can actually
solve the canonical-form condition for the two-body problem.

\subsection{\label{ssec:cntwo}The Conditions for the Two-body Case}

To solve the condition, we must first transform the variables from
$\sigma_{i}$ back to $h_{i}$. For this purpose it is convenient to
note that $h_{1}$ and $h_{2}$ are the solutions of the algebraic
equation:
\begin{align}
h_{i}^{2}-\sigma_{1}h_{i}+\sigma_{2}=0\qquad (i=1,2).
\label{eqn:alge2}
\end{align}
From the differential of the above relation:
\begin{align}
\dd h_{i}=\frac{h_{i}}{2h_{i}-\sigma_{1}}\dd\sigma_{1}
 -\frac{1}{2h_{i}-\sigma_{1}}\dd\sigma_{2}\qquad (i=1,2),
\label{eqn:diff2}
\end{align}
we can easily obtain,
\begin{align}
\frac{\partial}{\partial\sigma_{1}}=\sum_{i\neq j}^{2}
 \frac{h_{i}}{h_{i}-h_{j}}\frac{\partial}{\partial h_{i}},
\qquad
\frac{\partial}{\partial\sigma_{2}}=-\sum_{i\neq j}^{2}
 \frac{1}{h_{i}-h_{j}}\frac{\partial}{\partial h_{i}}.
\label{eqn:chai2}
\end{align}
Substituting Eqs.~(\ref{eqn:sigma}) and (\ref{eqn:chai2}) for
the gauged Hamiltonian (\ref{eqn:gsham}), we can express it in
terms of $h_{i}$ as follows:
\begin{align}
\tH_{\cN}
=&\,-\frac{1}{(h_{1}-h_{2})^{2}}\sum_{i=1}^{2}\bigl[\bA_{0}h_{i}^{4}
 -\bA_{1}h_{i}^{3}+(\bA_{2}+\bA_{11})h_{i}^{2}-\bA_{21}h_{i}
 +\bA_{22}\bigr]\frac{\partial^{2}}{\partial h_{i}^{2}}\notag\\
&\, +\frac{1}{(h_{1}-h_{2})^{2}}\bigl[2\bA_{0}\sigma_{2}^{2}
 -\bA_{1}\sigma_{2}\sigma_{1}+\bA_{2}\sigma_{1}^{2}
 -2(\bA_{2}-\bA_{11})\sigma_{2}\notag\\
&\, \qquad -\bA_{21}\sigma_{1}+2\bA_{22}\bigr]\left(
 \frac{\partial^{2}}{\partial h_{1}\partial h_{2}}
 +\sum_{i\neq j}^{2}\frac{1}{h_{i}-h_{j}}
 \frac{\partial}{\partial h_{i}}\right)\notag\\
&\, +\sum_{i\neq j}^{2}\frac{\bB_{0}h_{i}^{2}-\bB_{1}h_{i}
 +\bB_{2}}{h_{i}-h_{j}}\frac{\partial}{\partial h_{i}}-\bC.
\label{eqn:tHith}
\end{align}
Let us first consider the canonical-form condition with
respect to the second-order differential operators. Recalling
the relations:
\begin{subequations}
\label{eqns:smtfd}
\begin{gather}
\ee^{-\cW}\frac{\partial}{\partial h_{i}}\,\ee^{\cW}
=\frac{1}{h'_{i}}\left(\frac{\partial}{\partial q_{i}}
 +\frac{\partial\cW}{\partial q_{i}}\right),\\
\ee^{-\cW}\frac{\partial^{2}}{\partial h_{1}\partial h_{2}}\,\ee^{\cW}
=\frac{1}{h'_{1}h'_{2}}\left(\frac{
 \partial^{2}}{\partial q_{1}\partial q_{2}}+\sum_{i\neq j}^{2}
 \frac{\partial\cW}{\partial q_{i}}\frac{\partial}{\partial q_{j}}
 +\frac{\partial^{2}\cW}{\partial q_{1}\partial q_{2}}
 +\frac{\partial\cW}{\partial q_{1}}
 \frac{\partial\cW}{\partial q_{2}}\right),\\
\ee^{-\cW}\frac{\partial^{2}}{\partial h_{i}^{2}}\,\ee^{\cW}
=\frac{1}{(h'_{i})^{2}}\left[\frac{\partial^{2}}{\partial q_{i}^{2}}
 +\left(2\frac{\partial\cW}{\partial q_{i}}-\frac{h''_{i}}{h'_{i}}
 \right)\frac{\partial}{\partial q_{i}}
 +\frac{\partial^{2}\cW}{\partial q_{i}^{2}}+\left(
 \frac{\partial\cW}{\partial q_{i}}\right)^{2}-\frac{h''_{i}}{h'_{i}}
 \frac{\partial\cW}{\partial q_{i}}\right],
\end{gather}
\end{subequations}
and requiring the second-order differential operator to be
gauge-transformed back to the Laplacian in the flat space,
we obtain the following conditions:
\begin{align}
2\bA_{0}\sigma_{2}^{2}-\bA_{1}\sigma_{2}\sigma_{1}
 +\bA_{2}\sigma_{1}^{2}-2(\bA_{2}-\bA_{11})\sigma_{2}
 -\bA_{21}\sigma_{1}+2\bA_{22}=0,
\label{eqn:cfc1}\\
\bA_{0}h_{i}^{4}-\bA_{1}h_{i}^{3}+(\bA_{2}+\bA_{11})h_{i}^{2}
 -\bA_{21}h_{i}+\bA_{22}=\frac{1}{2}(h_{1}-h_{2})^{2}(h'_{i})^{2}.
\label{eqn:cfc2}
\end{align}
Once the above conditions satisfied, the gauged Hamiltonian is
transformed to,
\begin{align}
\ee^{-\cW}\tH_{\cN}\,\ee^{\cW}=&\,-\frac{1}{2}\sum_{i=1}^{2}\left[
 \frac{\del^{2}}{\del q_{i}^{2}}+\biggl(2\frac{\del\cW}{\del q_{i}}
 -\frac{h''_{i}}{h'_{i}}\biggr)\frac{\del}{\del q_{i}}
 +\frac{\del^{2}\cW}{\del q_{i}^{2}}+\biggl(\frac{\del\cW}{\del q_{i}}
 \biggr)^{2}-\frac{h''_{i}}{h'_{i}}\frac{\del\cW}{\del
 q_{i}}\right]\notag\\
&\,+\sum_{i\neq j}^{2}\frac{\bB_{0}h_{i}^{2}-\bB_{1}h_{i}+\bB_{2}}{
 (h_{i}-h_{j})h'_{i}}\biggl(\frac{\del}{\del q_{i}}
 +\frac{\del\cW}{\del q_{i}}\biggr)-\bC.
\label{eqn:stprH}
\end{align}
The condition that the above operator does not contain the first-order
differential operator reads,
\begin{align}
\frac{\partial\cW}{\partial q_{i}}=\frac{h''_{i}}{2h'_{i}}
 +\frac{\bB_{0}h_{i}^{2}-\bB_{1}h_{i}+\bB_{2}}{(h_{i}-h_{j})h'_{i}}
 \qquad (i\neq j).
\label{eqn:cfc3}
\end{align}
Finally, with the condition (\ref{eqn:cfc1}), (\ref{eqn:cfc2})
and (\ref{eqn:cfc3}) satisfied, the original
Hamiltonian takes the desirable Schr\"{o}dinger form as,
\begin{align}
H_{\cN}=\ee^{-\cW}\tH_{\cN}\,\ee^{\cW}
=-\frac{1}{2}\sum_{i=1}^{2}\frac{\partial^{2}}{\partial q_{i}^{2}}
 +\frac{1}{2}\sum_{i=1}^{2}\left[\left(
 \frac{\partial\cW}{\partial q_{i}}\right)^{2}
 -\frac{\partial^{2}\cW}{\partial q_{i}^{2}}\right]
 -\bC\bigl(\sigma(h)\bigr).
\label{eqn:res2H}
\end{align}
In the following, we shall solve the obtained canonical-form conditions
(\ref{eqn:cfc1}), (\ref{eqn:cfc2}) and (\ref{eqn:cfc3}) in order.

\subsection{\label{ssec:first}The first condition}

The first condition (\ref{eqn:cfc1}) is an algebraic identity
and thus is satisfied if and only if all the coefficients of the
polynomial vanish, i.e.,
\begin{eqnarray}
\begin{aligned}
A_{20,20}&=0, & 2A_{20,10}&=A_{20,21},
 & 2A_{10,10}&=A_{20,11}-A_{20,22},\\
A_{10,11}&=A_{20,12}, & A_{10,12}&=0,
 & A_{20,22}&=A_{21,21},\\
2A_{20,12}&=2A_{21,11}-A_{22,21}, & 2A_{10,12}&=2A_{11,11}-A_{22,11},
 & A_{12,11}&=0,\\
A_{21,01}&=-A_{22,22}, & A_{22,01}&=2A_{11,01}+2A_{22,12},
 & A_{12,01}&=2A_{12,12},\\
A_{01,01}&=-A_{22,02}, & A_{02,01}&=2A_{12,02},
 & A_{02,02}&=0.
\end{aligned}
\label{eqn:cfc11}
\end{eqnarray}
There are 27 parameters $A_{\kappa\lambda,\mu\nu}$ in
Eq.~(\ref{eqn:sgham}) for $M=2$. The above conditions reduce
the number of these free parameters to 12.

\subsection{\label{ssec:secon}The second condition}

With the first condition (\ref{eqn:cfc11}) satisfied, the l.h.s.
of the second condition (\ref{eqn:cfc2}) becomes,
\begin{align}
\lefteqn{
\bA_{0}h_{i}^{4}-\bA_{1}h_{i}^{3}+(\bA_{2}+\bA_{11})h_{i}^{2}
 -\bA_{21}h_{i}+\bA_{22}
}\notag\\
=&\,\left(A_{20,10}\sigma_{2}h_{i}^{4}+A_{10,10}\sigma_{1}h_{i}^{4}
 -A_{20,22}\sigma_{2}h_{i}^{3}+A_{21,11}\sigma_{2}h_{i}^{2}
 +A_{11,11}\sigma_{1}h_{i}^{2}\right.
\notag\\
&\, \left.-A_{22,22}\sigma_{2}h_{i}+A_{11,01}h_{i}^{2}
 -A_{22,12}\sigma_{2}-A_{22,02}h_{i}
 -A_{12,12}\sigma_{1}-A_{12,02}\right)(h_{i}-h_{j})\notag\\
&\, -A_{20,12}h_{i}^{3}(h_{i}+2h_{j})(h_{i}-h_{j})\qquad (i\neq j).
\label{eqn:cfc21}
\end{align}
The differential equation (\ref{eqn:cfc2}) has a solution if and
only if Eq.~(\ref{eqn:cfc21}) is divided by the factor
$(h_{1}-h_{2})^{2}$. This factorization condition reads,
\begin{eqnarray}
\begin{aligned}
A_{20,10}&=0, & A_{20,22}&=2A_{10,10}, & A_{21,11}&=3A_{20,12},\\
A_{22,22}&=2A_{11,11}, & A_{22,12}&=A_{11,01},
 & A_{22,02}&=-2A_{12,12}, & A_{12,02}&=0,
\end{aligned}
\label{eqn:cfc22}
\end{eqnarray}
from which the number of the free parameters $A_{\kappa\lambda,\mu\nu}$
is further reduced to 5. Then, the second condition turns to be,
\begin{align}
\frac{1}{2}(h'_{i})^{2}&=A_{10,10}h_{i}^{4}-A_{20,12}h_{i}^{3}
 +A_{11,11}h_{i}^{2}+A_{11,01}h_{i}+A_{12,12}\notag\\
&\equiv a_{4}h_{i}^{4}+a_{3}h_{i}^{3}+a_{2}h_{i}^{2}+a_{1}h_{i}+a_{0}
 \equiv P(h_{i}).
\label{eqn:cfc23}
\end{align}
Under the conditions (\ref{eqn:cfc11}) and (\ref{eqn:cfc22}),
the polynomials $\bA_{\kappa}$ and $\bA_{kl}$ are expressed in terms
of the five parameters $a_{p}$ $(p=0,\dots,4)$ as follows:
\begin{subequations}
\label{eqns:cfc2s}
\begin{align}
\bA_{0}(\sigma)&=a_{4}\sigma_{1}^{2},
\label{eqn:cfc2s1}\\
\bA_{1}(\sigma)&=4a_{4}\sigma_{2}\sigma_{1}-a_{3}\sigma_{1}^{2},
\label{eqn:cfc2s2}\\
\bA_{2}(\sigma)&=2a_{4}\sigma_{2}^{2}-a_{3}\sigma_{2}\sigma_{1},
\label{eqn:cfc2s3}\\
\bA_{11}(\sigma)&=2a_{4}\sigma_{2}^{2}-3a_{3}\sigma_{2}\sigma_{1}
 +a_{2}\sigma_{1}^{2}-2a_{2}\sigma_{2}+a_{1}\sigma_{1}+2a_{0},
\label{eqn:cfc2s4}\\
\bA_{21}(\sigma)&=-4a_{3}\sigma_{2}^{2}+2a_{2}\sigma_{2}\sigma_{1}
 +4a_{1}\sigma_{2}+2a_{0}\sigma_{1},
\label{eqn:cfc2s5}\\
\bA_{22}(\sigma)&=2a_{2}\sigma_{2}^{2}+a_{1}\sigma_{2}\sigma_{1}
 +a_{0}\sigma_{1}^{2}-2a_{0}\sigma_{2}.
\label{eqn:cfc2s6}
\end{align}
\end{subequations}

\subsection{\label{ssec:third}The third condition}

The third condition (\ref{eqn:cfc3}) is a set of partial differential
equations with two variables. In order that it has a solution,
we must impose the integrability condition:
\begin{align}
\frac{\partial}{\partial q_{2}}\frac{\partial\cW}{\partial q_{1}}
=\frac{\partial}{\partial q_{1}}\frac{\partial\cW}{\partial q_{2}}.
\label{eqn:slcd1}
\end{align}
Using Eq.~(\ref{eqn:cfc3}), we get,
\begin{align}
\frac{\partial}{\partial q_{j}}\frac{\partial\cW}{\partial q_{i}}
&=\frac{\partial}{\partial q_{j}}\left[\frac{h''_{i}}{h'_{i}}
 +\frac{\bB_{0}h_{i}^{2}-\bB_{1}h_{i}+\bB_{2}}{(h_{i}-h_{j})h'_{i}}
 \right]\notag\\
&=\frac{h'_{j}\sum_{p=0}^{4}\bar{a}_{p}h_{i}^{p}}%
{h'_{i}(h_{i}-h_{j})^{2}}\qquad (i\neq j),
\label{eqn:slcd2}
\end{align}
where $\bar{a}_{p}$ $(p=0,\dots,4)$ are given by,
\begin{eqnarray}
\begin{aligned}
\bar{a}_{0}&=-B_{02}, & \bar{a}_{1}&=B_{01}-2B_{12},\\
\bar{a}_{2}&=2B_{11}-B_{22},
 & \bar{a}_{3}&=(2\cN+3)a_{3}-2B_{10}+B_{21},\\
\bar{a}_{4}&=2(\cN-2)a_{4}-B_{20}.
\end{aligned}
\label{eqn:slcd3}
\end{eqnarray}
Therefore the integrability condition (\ref{eqn:slcd1}) is
equivalent to,
\begin{align}
\frac{\sum_{p=0}^{4}\bar{a}_{p}h_{1}^{p}}{\sum_{p=0}^{4}a_{p}
 h_{1}^{p}}=\frac{\sum_{p=0}^{4}\bar{a}_{p}h_{2}^{p}}{\sum_{p=0}^{4}
a_{p}h_{2}^{p}}.
\label{eqn:slcd4}
\end{align}
It is evident that the above relation holds if and only if both of
the quantities are a constant, $-2c$. Finally, the integrability
condition reads,
\begin{align}
\bar{a}_{p}=-2ca_{p}\qquad (p=0,\dots,4).
\label{eqn:slcd5}
\end{align}
Combining Eqs.~(\ref{eqn:slcd3}) and (\ref{eqn:slcd5}), we obtain,
\begin{eqnarray}
\begin{aligned}
B_{02}&=2c a_{0}, & B_{01}-2B_{12}&=-2c a_{1},\\
2B_{11}-B_{22}&=-2c a_{2}, & 2B_{10}-B_{21}&=(2\cN+3+2c)a_{3},\\
B_{20}&=2(\cN-2+c)a_{4},
\end{aligned}
\label{eqn:slcd6}
\end{eqnarray}
from which the number of the free parameters $B_{\kappa\lambda}$
reduces to 3. When the integrability condition is fulfilled,
the third condition (\ref{eqn:cfc3}) turns to be,
\begin{align}
\frac{\partial\cW}{\partial q_{i}}=\frac{h''_{i}}{2h'_{i}}
 +\frac{1}{h'_{i}}\left[\frac{\cN-2+c}{2}P'(h_{i})-Q(h_{i})\right]
 -\frac{ch'_{i}}{h_{i}-h_{j}}\qquad (i\neq j),
\label{eqn:cfc30}
\end{align}
where $Q(h_{i})$ is defined by,
\begin{align}
Q(h_{i})&\equiv\left(\frac{\cN-2-c}{2}a_{3}+B_{10}\right)h_{i}^{2}
 +\Bigl( (\cN-1-c)a_{2}-B_{11}\Bigr)h_{i}+\frac{\cN-2+c}{2}
 a_{1}-B_{12}\notag\\
&\equiv b_{2}h_{i}^{2}+b_{1}h_{i}+b_{0}.
\label{eqn:cfc31}
\end{align}
The parameters $B_{\kappa\lambda}$ are expressed in terms of the new
three parameters $b_{p}$ and the five $a_{p}$ as,
\begin{subequations}
\label{eqns:cfc32}
\begin{align}
B_{10}&=b_{2}-\frac{\cN-2-c}{2}a_{3},
 & B_{20}&=2(\cN-2+c)a_{4},
\label{eqn:cfc321}\\
B_{11}&=-b_{1}+(\cN-1-c)a_{2},
 & B_{21}&=2b_{2}-(3\cN+1+c)a_{3},
\label{eqn:cfc322}\\
B_{12}&=-b_{0}+\frac{\cN-2+c}{2}a_{1},
 & B_{22}&=-2b_{1}+2(\cN-1)a_{2},
\label{eqn:cfc323}\\
B_{01}&=-2b_{0}+(\cN-2-c)a_{1},
 & B_{02}&=2c a_{0}.
\label{eqn:cfc324}
\end{align}
\end{subequations}
We now can easily integrate the differential equation (\ref{eqn:cfc30})
to obtain,
\begin{align}
\cW(q_{1},q_{2})=-\sum_{i=1}^{2}\int\dd h_{i}\frac{Q(h_{i})}{2P(h_{i})}
+\frac{\cN-1+c}{2}\sum_{i=1}^{2}\ln |h'_{i}|-c\ln |h_{1}-h_{2}|,
\label{eqn:cfc33}
\end{align}
where the integral constant is omitted.

\subsection{\label{ssec:summ4}Summary}

In summary, we have derived the canonical-form conditions for
two-body case, Eqs.~(\ref{eqn:cfc1}), (\ref{eqn:cfc2}) and
(\ref{eqn:cfc3}), and have solved them to obtain
Eqs.~(\ref{eqn:cfc23}), (\ref{eqns:cfc2s}) and
(\ref{eqn:cfc31})--(\ref{eqn:cfc33}). 
Substituting Eqs.~(\ref{eqn:cfc1}), (\ref{eqn:cfc2}), (\ref{eqn:cfc3})
and (\ref{eqn:cfc30}) for Eq.~(\ref{eqn:tHith}), we see that
the gauged Hamiltonian $\tH_{\cN}$ which satisfies the
canonical-form condition must have the following form:
\begin{align}
\tH_{\cN}(h)=&\,-\sum_{i=1}^{2}P(h_{i})\frac{\partial^{2}}{
 \partial h_{i}^{2}}
 +\sum_{i=1}^{2}\left[\frac{\cN-2+c}{2}P'(h_{i})-Q(h_{i})\right]
 \frac{\partial}{\partial h_{i}}\notag\\
&\, -2c\sum_{i\neq j}^{2}\frac{P(h_{i})}{h_{i}-h_{j}}
 \frac{\partial}{\partial h_{i}}-\bC\bigl(\sigma(h)\bigr).
\label{eqn:cfc34}
\end{align}
In the above, $\bC$ is calculated by substituting
Eqs.~(\ref{eqn:cfc2s1}) and (\ref{eqn:cfc321}) for
Eq.~(\ref{eqn:sopol5}) and reads,
\begin{align}
\bC\bigl(\sigma(h)\bigr)=&\,
 \frac{\cN -1}{12}(\cN -2+2c)\sum_{i=1}^{2}P''(h_{i})
\notag\\
&\, -\frac{\cN -1}{2}\sum_{i=1}^{2}Q'(h_{i})
-\frac{\cN -1}{2}c\sum_{i\neq j}^{2}\frac{P'(h_{i})}{h_{i}-h_{j}}+R,
\label{eqn:cfc35}
\end{align}
where the constant $R$ is given by,
\begin{align}
R=-\frac{(\cN-1)(\cN-2-c)}{3}a_{2}+(\cN-1)b_{1}+C.
\label{eqn:cfc36}
\end{align}

\section{\label{sec:gl2ks}$GL(2,K)$ Shape Invariance}

It was shown that the one-body $\fsl (2)$ quasi-solvable models can
be classified using the shape invariance of the Hamiltonian
under the action of $GL(2,K)$ ($K=\bbR$ or $\bbC$) of linear fractional
transformations~\cite{LoKaOl3,LoKaOl4}. 
We can see that the two-body
Hamiltonian (\ref{eqn:cfc34}) also has the same property of shape
invariance. The linear fractional transformation of $h_{i}$ is
introduced by,
\begin{align}
h_{i}\mapsto\hat{h}_{i}=\frac{\alpha h_{i}+\beta}{\gamma h_{i}+\delta}
\qquad (\alpha,\beta,\gamma,\delta\in K;\ \Delta\equiv\alpha\delta
 -\beta\gamma\neq 0).
\label{eqn:frtsf}
\end{align}
Then, it turns out that the Hamiltonian (\ref{eqn:cfc34}) is
shape invariant under the following transformation (with $M=2$)
induced by Eq.~(\ref{eqn:frtsf}),
\begin{align}
\tH_{\cN}(h)\mapsto\widehat{\tH}_{\cN}(h)=
\prod_{i=1}^{M}(\gamma h_{i}+\delta)^{\cN -1}\,\tH_{\cN}
(\hat{h})\prod_{i=1}^{M}(\gamma h_{i}+\delta)^{-(\cN -1)},
\label{eqn:tfham}
\end{align}
where the polynomials $P(h_{i})$ and $Q(h_{i})$ in the
$\tH_{\cN}(h)$ are transformed according to,
\begin{subequations}
\label{eqns:tfpol}
\begin{align}
P(h_{i})&\mapsto\hat{P}(h_{i})=\Delta^{-2}(\gamma h_{i}+\delta)^{4}
 P(\hat{h}_{i}),\label{eqn:tfofP}\\
Q(h_{i})&\mapsto\hat{Q}(h_{i})=\Delta^{-1}(\gamma h_{i}+\delta)^{2}
 Q(\hat{h}_{i}).\label{eqn:tfofQ}
\end{align}
\end{subequations}
This shape invariance originally comes from the $GL(2,K)$
invariance of the solvable subspace (\ref{eqn:svspc}):
\begin{align}
\tilde{\cV}_{\cN}(h)\mapsto\widehat{\tilde{\cV}}_{\cN}(h)
&\equiv\mathrm{span}\,\left\{\prod_{i=1}^{M}(\gamma h_{i}
 +\delta)^{\cN-1}\tilde{\phi}: \tilde{\phi}\in\tilde{\cV}_{\cN}
 (\hat{h})\right\}
 \notag\\
&=\tilde{\cV}_{\cN}(h),
\label{eqn:invsp}
\end{align}
where $\tilde{\cV}_{\cN}(\hat{h})$ is given by,
\begin{align}
\tilde{\cV}_{\cN}(\hat{h})=\mathrm{span}\,\left\{\sigma_{1}
 (\hat{h})^{n_{1}}\cdots\sigma_{M}(\hat{h})^{n_{M}}:
 n_{i}\in\bbZ_{\ge 0},\, 0\le\sum_{i=1}^{M}n_{i}\le\cN -1\right\}.
\label{eqn:trssp}
\end{align}
From the invariance (\ref{eqn:invsp}), the gauged Hamiltonian
(\ref{eqn:sgham}) must be invariant under the $GL(2,K)$
transformation (\ref{eqn:tfham}) because it is constructed
so that $\tH_{\cN}\tilde{\cV}_{\cN}\subset\tilde{\cV}_{\cN}$.
Therefore, the gauged Hamiltonian (\ref{eqn:cfc34}) constitutes
an invariant subspace of the $GL(2,K)$ invariant operators for
$M=2$ which satisfy the canonical-form condition.

From the above consideration, the $M$-body extension $(M>2)$
of the two-body gauged Hamiltonian (\ref{eqn:cfc34}) should
have the same shape invariance property under the $GL(2,K)$
transformation (\ref{eqn:tfham}). From the functional form
of the two-body $\fsl(3)$ quasi-solvable system (\ref{eqn:cfc34})
and also from the form of the one-body $\fsl(2)$ quasi-solvable
system~\cite{Turbi1,LoKaOl3,LoKaOl4,Tanak3}, we can cast the $M$-body
quasi-solvable gauged Hamiltonian which contains up to two-body
interactions as,
\begin{align}
\tH_{\cN}(h)
=&\,-\sum_{i=1}^{M}P(h_{i})\frac{\partial^{2}}{\partial h_{i}^{2}}
 +\sum_{i=1}^{M}\bigl[C_{1}(M)P'(h_{i})-Q(h_{i})\bigr]
 \frac{\partial}{\partial h_{i}}\notag\\
&\, -2c\sum_{i\neq j}^{M}\frac{P(h_{i})}{h_{i}-h_{j}}
 \frac{\partial}{\partial h_{i}}-\left[ \frac{\cN-1}{12}
 C_{2}(M)\sum_{i=1}^{M}P''(h_{i})\right.\notag\\
&\,\left.-\frac{\cN-1}{2}\sum_{i=1}^{M}Q'(h_{i})
 -\frac{\cN -1}{2}C_{3}(M)\sum_{i\neq j}^{M}
 \frac{P'(h_{i})}{h_{i}-h_{j}}+R \right],
\label{eqn:azham}
\end{align}
where $C_{1}$, $C_{2}$ and $C_{3}$ satisfy the following initial
conditions:
\begin{subequations}
\label{eqns:3Cicds}
\begin{alignat}{3}
C_{1}(1)&=\frac{\cN-2}{2}, & \qquad C_{1}(2)&=\frac{\cN-2+c}{2},\\
C_{2}(1)&=\cN-2, & C_{2}(2)&=\cN-2+2c,\\
C_{3}(2)&=c.
\end{alignat}
\end{subequations}
If we make the $GL(2,K)$ transformation on the above $\tH_{\cN}$
according to Eq.~(\ref{eqn:tfham}), we obtain,
\begin{align}
\widehat{\tH}_{\cN}(h)
=&\,-\sum_{i=1}^{M}\hat{P}(h_{i})\frac{\partial^{2}}{\partial h_{i}^{2}}
 +\sum_{i=1}^{M}\bigl[C_{1}(M)P_{1}(h_{i})-\hat{Q}(h_{i})\bigr]
 \frac{\partial}{\partial h_{i}}\notag\\
&\, -2c\sum_{i\neq j}^{M}\frac{\hat{P}(h_{i})}{h_{i}-h_{j}}
 \frac{\partial}{\partial h_{i}}-\left[ \frac{\cN -1}{12}C_{2}(M)
 \sum_{i=1}^{M}P_{2}(h_{i})\right.\notag\\
&\, \left.-\frac{\cN-1}{2}\sum_{i=1}^{M}\hat{Q}'(h_{i})
 -\frac{\cN -1}{2}C_{3}(M)\sum_{i\neq j}^{M}
 \frac{P_{3}(h_{i})}{h_{i}-h_{j}}+R\right],
\label{eqn:Haztf}
\end{align}
where $P_{1}$, $P_{2}$ and $P_{3}$ are given by,
\begin{subequations}
\label{eqns:tfcofs}
\begin{align}
P_{1}(h_{i})=&\,\Delta^{-1}(\gamma h_{i}+\delta)^{2}
 P'(\hat{h}_{i})+2\,\frac{\cN-2+(M-1)c}{C_{1}(M)}\Delta^{-2}
 \gamma (\gamma h_{i}+\delta)^{3}P(\hat{h}_{i}),\\
P_{2}(h_{i})=&\, P''(\hat{h}_{i})+6\,
 \frac{2C_{1}(M)+(M-1)C_{3}(M)}{C_{2}(M)}\Delta^{-1}\gamma
 (\gamma h_{i}+\delta)P'(\hat{h}_{i})\notag\\
&\, +12\,\frac{\cN-2+2(M-1)c}{C_{2}(M)}\Delta^{-2}\gamma^{2}
 (\gamma h_{i}+\delta)^{2}P(\hat{h}_{i}),\\
P_{3}(h_{i})=&\,\Delta^{-1}(\gamma h_{i}+\delta)^{2}
 P'(\hat{h}_{i})+\frac{4c}{C_{3}(M)}\Delta^{-2}\gamma
 (\gamma h_{i}+\delta)^{3}P(\hat{h}_{i}).
\end{align}
\end{subequations}
Useful formulae needed in the above calculation are summarized in
Appendix \ref{sec:fmlGL}. Hence, the requirement that the $M$-body
gauged Hamiltonian is shape invariant under Eqs.~(\ref{eqn:tfham})
and (\ref{eqns:tfpol}) reads,
\begin{align}
P_{1}(h_{i})=P_{3}(h_{i})=\hat{P}'(h_{i}),\qquad
 P_{2}(h_{i})=\hat{P}''(h_{i}).
\label{eqn:rqinv}
\end{align}
Comparing Eqs.~(\ref{eqns:tfcofs}) with (\ref{eqns:dvptfs}),
we obtain conditions for the invariance:
\begin{align}
C_{1}(M)=\frac{\cN-2+(M-1)c}{2},\quad C_{2}(M)=\cN-2+2(M-1)c,
 \quad C_{3}(M)=c,
\label{eqn:invcd}
\end{align}
which in turn satisfy the required initial conditions
(\ref{eqns:3Cicds}). Thus, the requirement of the $GL(2,K)$
invariance uniquely determines the $M$-dependence of the coupling
constants as was mentioned before. Finally, we obtain,
\begin{align}
\tH_{\cN}(h)=&\,-\sum_{i=1}^{M}P(h_{i})\frac{\partial^{2}}{
 \partial h_{i}^{2}}
 +\sum_{i=1}^{M}\left[\frac{\cN-2+(M-1)c}{2}P'(h_{i})-Q(h_{i})\right]
 \frac{\partial}{\partial h_{i}}\notag\\
&\, -2c\sum_{i\neq j}^{M}\frac{P(h_{i})}{h_{i}-h_{j}}
 \frac{\partial}{\partial h_{i}}-\bC\bigl(\sigma(h)\bigr),
\label{eqn:htham}
\end{align}
where $\bC$ is given by,
\begin{align}
\bC\bigl(\sigma(h)\bigr)=&\,
\frac{\cN -1}{12}\bigl[\cN -2+2(M-1)c\bigr]\sum_{i=1}^{M}P''(h_{i})
\notag\\
&\, -\frac{\cN -1}{2}\sum_{i=1}^{M}Q'(h_{i})
-\frac{\cN -1}{2}c\sum_{i\neq j}^{M}\frac{P'(h_{i})}{h_{i}-h_{j}}+R.
\label{eqn:bCofh}
\end{align}
The $P$ and $Q$ in Eqs.~(\ref{eqn:htham}) and (\ref{eqn:bCofh})
are again a fourth- and a second-degree polynomial, respectively:
\begin{subequations}
\label{eqns:defpol}
\begin{align}
P(h_{i})&=a_{4}h_{i}^{4}+a_{3}h_{i}^{3}+a_{2}h_{i}^{2}+a_{1}h_{i}+a_{0},
\label{eqn:defP}\\
Q(h_{i})&=b_{2}h_{i}^{2}+b_{1}h_{i}+b_{0}.
\label{eqn:defQ}
\end{align}
\end{subequations}
There are 10 parameters, namely, $a_{p}$ $(p=0,\dots,4)$, $b_{p}$
$(p=0,1,2)$, $c$, $R$, which characterize the quasi-solvable
Hamiltonian (\ref{eqn:htham}).
The function $h(q)$, which determine the change of variables,
is given by a solution of the differential equation,
\begin{align}
h'(q)^{2}=2P\bigl(h(q)\bigr).
\label{eqn:dhofq}
\end{align}
If we transform back the gauged Hamiltonian (\ref{eqn:htham})
with $h(q)$ satisfying Eq.~(\ref{eqn:dhofq}), the original
Hamiltonian becomes the Schr\"{o}dinger type,
\begin{align}
H_{\cN}=-\frac{1}{2}\sum_{i=1}^{M}\frac{\partial^{2}}{
 \partial q_{i}^{2}}+\frac{1}{2}\sum_{i=1}^{M}\left[\left(
 \frac{\partial\cW(q)}{\partial q_{i}}\right)^{2}
 -\frac{\partial^{2}\cW(q)}{\partial q_{i}^{2}}\right]
 -\bC\bigl(\sigma(h)\bigr),
\label{eqn:orham}
\end{align}
and the \textit{superpotential} $\cW(q)$ is given by,
\begin{align}
\cW(q)=-\sum_{i=1}^{M}\int\dd h_{i}\frac{Q(h_{i})}{2P(h_{i})}
 +\frac{\cN -1+(M-1)c}{2}\sum_{i=1}^{M}\ln\left| h'_{i}\right|
 -c\sum_{i<j}^{M}\ln\left| h_{i}-h_{j}\right|.
\label{eqn:sppot}
\end{align}
By construction, the elements of the vector space (\ref{eqn:svosp})
give the solvable wave functions $\psi(q)$ of the Hamiltonian
(\ref{eqn:orham}):
\begin{align}
\psi(q)\in\mathrm{span}\,\left\{\sigma_{1}^{n_{1}}\cdots
 \sigma_{M}^{n_{M}}\ee^{-\cW(q)}: n_{i}\in\bbZ_{\ge 0},\,
 0\le\sum_{i=1}^{M}n_{i}\le\cN -1\right\}.
\label{eqn:svwav}
\end{align}

From the form of the Hamiltonian (\ref{eqn:orham}), we can observe
an interesting feature of the (quasi-)solvability.
As we have examined in the end of Section~\ref{sec:algeb}, the system
is solvable if Eq.~(\ref{eqn:solc21}) is fulfilled. The resulting
equation (\ref{eqn:solc23}) together with the form of the Hamiltonian
(\ref{eqn:orham}) indicate that the system becomes of supersymmetric
form (except for the irrelevant constant term $-C$) \cite{Witte1,Witte2}
if it is not only quasi-solvable but also solvable.
A system is always quasi-solvable if it is supersymmetric,
since the ground state is always solvable.
From the above result, we can conclude that a system is always
supersymmetric if it is solvable and all its states have the form
(\ref{eqn:svwav}). We note that this interesting feature does not
depend on the normalizability of the solvable sector, that is,
it holds whether or not supersymmetry is dynamically broken.

\section{\label{sec:inver}Inverse Problem}

In the preceding section, we have derived the $M$-body Hamiltonian
containing up to the two-body interactions (\ref{eqn:htham}),
without directly solving the canonical-form condition for $M>2$.
To confirm that the Hamiltonian (\ref{eqn:htham}) is indeed
quasi-solvable, i.e., it preserves the vector space (\ref{eqn:svspc})
for any integer $M$, we shall see in this section that it can be surely
written in terms of the elementary symmetric polynomials $\sigma_{i}$
as the form of Eq.~(\ref{eqn:sgham}) with the set of the specific form
of the polynomials (\ref{eqns:sopols}). In order to express the
Hamiltonian (\ref{eqn:htham}) in terms of $\sigma_{i}$, we need
to know the following transformation laws from $h_{i}$ to $\sigma_{i}$:
\begin{subequations}
\label{eqns:tflaws}
\begin{align}
\sum_{i=1}^{M}h_{i}^{p}\frac{\partial^{2}}{\partial h_{i}^{2}}
 &=\sum_{k,l=1}^{M}A_{k,l}^{(p)}(\sigma)\frac{\partial^{2}}{\partial
 \sigma_{k}\partial\sigma_{l}} \qquad (p=0,\dots,4),\\
\sum_{i=1}^{M}h_{i}^{p}\frac{\partial}{\partial h_{i}}
 &=\sum_{k=1}^{M}B_{k}^{(p)}(\sigma)\frac{\partial}{\partial\sigma_{k}}
 \qquad (p=0,\dots,3),\\
\sum_{i=1}^{M}h_{i}^{p}&=B_{-}^{(p)}(\sigma) \qquad (p=0,1,2),\\
\sum_{i\neq j}^{M}\frac{h_{i}^{p}}{h_{i}-h_{j}}\frac{\partial}{
 \partial h_{i}}
 &=\sum_{k=1}^{M}C_{k}^{(p)}(\sigma)\frac{\partial}{\partial\sigma_{k}}
 \qquad (p=0,\dots,4),\\
\sum_{i\neq j}^{M}\frac{h_{i}^{p}}{h_{i}-h_{j}}
 &=C_{-}^{(p)}(\sigma) \qquad (p=0,\dots,3).
\end{align}
\end{subequations}
Substituting the above for Eqs.~(\ref{eqn:htham}) and
(\ref{eqn:bCofh}), we rewrite the gauged Hamiltonian as,
\begin{align}
\tH_{\cN}
=&\,-\sum_{k,l=1}^{M}\left(\sum_{p=0}^{4}a_{p} A_{k,l}^{(p)}\right)
 \frac{\partial^{2}}{\partial\sigma_{k}\partial\sigma_{l}}
 -\sum_{k=1}^{M}\Biggl[\sum_{p=0}^{2}b_{p}B_{k}^{(p)}\notag\\
&\, -\frac{\cN-2+(M-1)c}{2}\sum_{p=0}^{3}(p+1)a_{p+1}B_{k}^{(p)}
 +2c\sum_{p=0}^{4}a_{p}C_{k}^{(p)}\Biggr]
 \frac{\partial}{\partial\sigma_{k}}
 -\bC(\sigma),
\label{eqn:cvham}
\end{align}
with,
\begin{align}
\bC(\sigma)
=&\,\frac{\cN-1}{12}\bigl[\cN-2+2(M-1)c\bigr]\sum_{p=0}^{2}
 (p+2)(p+1)a_{p+2}B_{-}^{(p)}\notag\\
&\, -\frac{\cN-1}{2}\sum_{p=0}^{1}(p+1)b_{p+1}B_{-}^{(p)}
 -\frac{\cN-1}{2}c\sum_{p=0}^{3}(p+1)a_{p+1}C_{-}^{(p)}+R.
\label{eqn:cvbC1}
\end{align}
In Appendix~\ref{sec:tflIP}, we give the explicit procedure to obtain
the transformation formulae (\ref{eqns:tflaws}).

Let us begin with investigating $\bC(\sigma)$. Applying the
formulae (\ref{eqns:formB-}) and (\ref{eqns:formC-}) to
Eq.~(\ref{eqn:cvbC1}), we obtain,
\begin{align}
\bC(\sigma)
=&\,(\cN-1)(\cN-2)a_{4}\sigma_{1}^{2}-2(\cN-1)(\cN-2+c)
 a_{4}\sigma_{2}\notag\\
&\, -(\cN-1)\left(b_{2}-\frac{\cN-2-(M-1)c}{2}a_{3}\right)
 \sigma_{1}\notag\\
&\, +(\cN-1)\frac{\cN-2-(M-1)c}{6}Ma_{2}-\frac{\cN-1}{2}Mb_{1}+R.
\label{eqn:cvbC2}
\end{align}
Comparing the result (\ref{eqn:cvbC2}) with Eq.~(\ref{eqn:sopol5}),
we can easily see that the $\bC(\sigma)$ in the Hamiltonian
(\ref{eqn:htham}) is indeed of the form (\ref{eqn:sopol5}) with,
\begin{align}
\bA_{0}(\sigma)&=a_{4}\sigma_{1}^{2},
\label{eqn:cvfm1}\\
\sum_{i=1}^{M}B_{i0}\sigma_{i}&=2(\cN-2+c)a_{4}\sigma_{2}
 +\left(b_{2}-\frac{\cN-2-(M-1)c}{2}a_{3}\right)\sigma_{1},
\label{eqn:cvfm2}\\
C&=(\cN-1)\frac{\cN-2-(M-1)c}{6}Ma_{2}-\frac{\cN-1}{2}Mb_{1}+R.
\label{eqn:cvfm3}
\end{align}
These results are in agreement, in the case of $M=2$, with
those obtained in Section~\ref{sec:canon}, namely,
Eqs.~(\ref{eqn:cfc2s1}), (\ref{eqn:cfc321}) and (\ref{eqn:cfc36}).

In the next, we will examine the structure of the second-order
derivative operator. Applying the formulae (\ref{eqns:formuA})
to the first term of Eq.~(\ref{eqn:cvham}), we obtain,
\begin{align}
\sum_{p=0}^{4}a_{p}A_{k,l}^{(p)}
=&\, a_{4}\sigma_{1}^{2}\sigma_{k}\sigma_{l}
 -\bigl[(2a_{4}\sigma_{2}-a_{3}\sigma_{1})\sigma_{k}+a_{4}\sigma_{1}
 \sigma_{k+1}\bigr]\sigma_{l}\notag\\
&\, -a_{4}\sigma_{1}\sigma_{k}\sigma_{l+1}+A_{kl}(\sigma),
\label{eqn:cvfm4}
\end{align}
where $A_{kl}(\sigma)$ represents the part involving up to the
quadratic terms in $\sigma_{i}$ and is given by,
\begin{align}
\lefteqn{
A_{kl}(\sigma)
=a_{4}\left[\sum_{m=0}^{2}(k-m+2)\sigma_{k-m+2}\sigma_{l+m}
 +\sum_{m=3}^{k+2}(k-l-2m+2)\sigma_{k-m+2}\sigma_{l+m}\right]
}\notag\\
&{}-a_{3}\left[\sum_{m=0}^{1}(k-m+1)\sigma_{k-m+1}\sigma_{l+m}
 +\sum_{m=2}^{k+1}(k-l-2m+1)\sigma_{k-m+1}\sigma_{l+m}\right]\notag\\
&{}+a_{2}\left[k\sigma_{k}\sigma_{l}+\sum_{m=1}^{k}(k-l-2m)
 \sigma_{k-m}\sigma_{l+m}\right]-a_{1}\sum_{m=0}^{k-1}
 (k-l-2m-1)\sigma_{k-m-1}\sigma_{l+m}\notag\\
&{}+a_{0}\left[(M-l+1)\sigma_{k-1}\sigma_{l-1}+\sum_{m=0}^{k-2}
 (k-l-2m-2)\sigma_{k-m-2}\sigma_{l+m}\right].
\label{eqn:cvfm5}
\end{align}
In Eq.~(\ref{eqn:cvfm5}), we note that $\sigma_{0}=1$ and
$\sigma_{-k}=0$ for $k>0$. Comparing Eq.~(\ref{eqn:cvfm4}) with
Eq.~(\ref{eqn:sgham}) and (\ref{eqn:sopol1})--(\ref{eqn:sopol3}),
we can observe that
the second-order derivative operator in Eq.~(\ref{eqn:htham})
are expressed in the form of the first line in Eq.~(\ref{eqn:sgham})
with the substitution (\ref{eqn:cvfm1}) and
\begin{align}
\bA_{k}(\sigma)&=(2a_{4}\sigma_{2}-a_{3}\sigma_{1})\sigma_{k}
 +2a_{4}\sigma_{1}\sigma_{k+1},
\label{eqn:cvfm6}\\
\bA_{kk}(\sigma)&=A_{kk}(\sigma),
\label{eqn:cvfm7}\\
\bA_{kl}(\sigma)&=A_{kl}(\sigma)+A_{lk}(\sigma)\qquad (k>l).
\label{eqn:cvfm8}
\end{align}
In the case of $M=2$, these results are also in agreement with
Eqs.~(\ref{eqns:cfc2s}) obtained in Section~\ref{sec:canon}.

Finally, if we apply the formulae (\ref{eqns:formuB}) and
(\ref{eqns:formuC}) to the second term of Eq.~(\ref{eqn:cvham}),
we can find that the first-order operator in the gauged Hamiltonian
(\ref{eqn:htham}) has the form of the first term in the second line
of Eq.~(\ref{eqn:sgham}) with the following relation:
\begin{align}
\sum_{i\geq j(\geq k)}^{M}A_{ij,jk}\sigma_{i}&-\sum_{i=1}^{M}B_{ik}
 \sigma_{i}-B_{0k}=-(k+2)\bigl[2(\cN-2)+(k+1)c\bigr]a_{4}\sigma_{k+2}
 \notag\\
&-(k+1)\biggl(b_{2}-\frac{3(\cN-2)+(2k-M+1)c}{2}a_{3}\biggr)\sigma_{k+1}
 \notag\\
&+k\Bigl(b_{1}-\bigl[\cN-2+(k-M)c\bigr]a_{2}\Bigr)\sigma_{k}\notag\\
&-(k-M-1)\biggl(b_{0}-\frac{\cN-2+(2k-M-1)c}{2}a_{1}\biggr)\sigma_{k-1}
 \notag\\
&-(k-M-1)(k-M-2)ca_{0}\sigma_{k-2}.
\label{eqn:cvfm9}
\end{align}
From this relation, we can determine $B_{ik}$ and $B_{0k}$ uniquely
since the coefficients $A_{ij,jk}$ in the l.h.s. of
Eq.~(\ref{eqn:cvfm9}) are fixed by Eqs.~(\ref{eqn:sopol2}) and
(\ref{eqn:cvfm7})--(\ref{eqn:cvfm8}). In the case of $M=2$, we can
again check that $B_{ik}$ and $B_{0k}$ calculated by
Eq.~(\ref{eqn:cvfm9}) are in complete agreement with
Eqs.~(\ref{eqn:cfc322})--(\ref{eqn:cfc324}) derived in
Section~\ref{sec:canon}.

Summarizing the analyses, we have shown that the $M$-body gauged
Hamiltonian (\ref{eqn:htham}) can be written in terms of
$\sigma_{i}$ as the form of Eq.~(\ref{eqn:sgham}), which ensures
the quasi-solvability with the solvable sector (\ref{eqn:svspc}).
In addition, we can find out a reason why the form of the quasi-solvable
gauged Hamiltonian should be restricted to Eq.~(\ref{eqn:htham}).
For example, the formulae (\ref{eqns:formB-}) and (\ref{eqn:form31})
indicate,
\begin{align}
\sum_{i=1}^{M}h_{i}^{p}\frac{\partial}{\partial h_{i}}
 \sim\sum_{k=1}^{M}\sigma_{1}^{p-1}\sigma_{k}\frac{\partial}{
 \partial\sigma_{k}}+\cdots.
\label{eqn:asBkp}
\end{align}
From the fact that $\bB_{0}(\sigma)$ in Eq.~(\ref{eqn:sgham}) is a
polynomial of at most second-degree, the highest degree term in the
r.h.s. of Eq.~(\ref{eqn:asBkp}) can be fit in the form of
Eq.~(\ref{eqn:sgham}) only if $p\le 3$. This explains why the
coefficient of the first-order derivative in Eq.~(\ref{eqn:htham})
must be a polynomial of at most third-degree in $h_{i}$. Similar
observation on the other parts of the operators can also explain
the structure of Eq.~(\ref{eqn:htham}).

Finally, we will return to the condition for the solvability.
As we have examined in Section~\ref{sec:algeb}, the operator
(\ref{eqn:sgham}) becomes solvable when the condition (\ref{eqn:solc21})
is satisfied. Comparing Eq.~(\ref{eqn:solc21}) with
Eqs.~(\ref{eqn:cvfm1}), (\ref{eqn:cvfm2}) and (\ref{eqn:cvfm6}),
we obtain the solvability condition in terms of the free parameters:
\begin{align}
a_{3}=a_{4}=b_{2}=0.
\label{eqn:slcfp}
\end{align}
Since the system has the $GL(2,K)$ shape invariance (\ref{eqn:tfham}),
it is solvable if the polynomials $P$ and $Q$ can be transformed
according to Eqs.~(\ref{eqns:tfpol}) so that the conditions
(\ref{eqn:slcfp}) are satisfied. In the next section, we will see
what kind of solvable models as well as quasi-solvable models can
be constructed from the general form of the Hamiltonian
(\ref{eqn:orham}).

\section{\label{sec:class}Classification of the Models}

For a given $P(h)$, the function $h(q)$ is determined by
Eq.~(\ref{eqn:dhofq}) and a particular model is obtained by
substituting this $h(q)$ for Eqs.~(\ref{eqn:bCofh}),
(\ref{eqn:orham}) and (\ref{eqn:sppot}). The solution of
Eq.~(\ref{eqn:dhofq}) is easily obtained as an elliptic integral:
\begin{align}
|q|=\int\frac{\dd h}{\sqrt{2P(h)}}.
\label{eqn:solqh}
\end{align}
We note that the above solution has the reflection and
translational symmetry; if $h(q)$ is a solution of Eq.~(\ref{eqn:dhofq}),
$h(-q)$ and $h(q+q_{0})$ also satisfy Eq.~(\ref{eqn:dhofq}).
Furthermore, this elliptic integral also has $GL(2,K)$ invariance
discussed in Section~\ref{sec:gl2ks}:
\begin{align}
\int\frac{\dd h}{\sqrt{2P(h)}}
\mapsto\int\frac{\dd h}{\sqrt{2\hat{P}(h)}}
&=\int\dd h\frac{\Delta}{(\gamma h+\delta)^{2}}
 \frac{1}{\sqrt{2P(\hat{h})}}\notag\\
&=\int\frac{\dd\hat{h}}{\sqrt{2P(\hat{h})}}.
\label{eqn:tfqoh}
\end{align}
The elliptic integral (\ref{eqn:solqh}) is classified according
to the distribution of the zeros of $P(h)$, e.g., multiplicity of
the zeros. Since the transformation (\ref{eqn:tfofP}) induced
by Eq.~(\ref{eqn:frtsf}) does not alter the distribution,
the shape invariance (\ref{eqn:tfqoh}) together with (\ref{eqn:tfham})
enable us to classify the resultant quasi-solvable models
(\ref{eqn:orham}) according to the distribution of the roots of $P(h)$.
This idea was first introduced in Ref.~\cite{LoKaOl3} to classify
the one-body $\fsl(2)$ quasi-solvable models.
Under the transformation (\ref{eqn:tfofP}) of $GL(2,\bbR)$ or
$GL(2,\bbC)$, every quartic polynomial $P(h)$ with real or complex
coefficients is equivalent to one of the eight or five forms,
respectively, shown in Table~\ref{tab:class}.
\begin{table}[h]
\begin{center}
\[
\arraycolsep=5mm
\begin{array}{|c|c|c|}\hline
\text{Case} & GL(2,\bbR) & GL(2,\bbC) \\
\hline
\text{I} & 1/2 &1/2 \\
\hline
\text{II} & 2h & 2h \\
\hline
\text{III} & 2\nu h^{2} & 2\nu h^{2} \\
\text{III$'$} & \nu (h^{2}+1)^{2}/2 & \\
\hline
\text{IV} & 2\nu (h^{2}-1) & 2\nu (h^{2}-1) \\
\text{IV$'$} & 2\nu(h^{2}+1) & \\
\hline
\text{V} & 2h^{3}-g_{2}h/2-g_{3}/2  & 2h^{3}-g_{2}h/2-g_{3}/2 \\
\text{V$'$} & \nu (h^{2}+1)[(1-k^{2})h^{2}+1]/2 & \\
\hline
\end{array}
\]
\caption{The representatives of $P(h)$ under the $GL(2,\bbR)$ and
$GL(2,\bbC)$ transformations.}
\label{tab:class}
\end{center}
\end{table}
In Table~\ref{tab:class}, $\nu,g_{2},g_{3}\in K$ according to the
transformation group $GL(2,K)$, and $\nu\neq 0$, $0<k<1$,
$g_{2}^{3}-27g_{3}^{2}\neq 0$.
Since case III$'$ under the $GL(2,\bbR)$ can be obtained from
case III by a $GL(2,\bbC)$ transformation, the potential in case
III$'$ is regarded as another real function representation of the
complex potential \textit{real-valued on} $\bbR^{M}$ in case III.
The same relation holds between case IV and IV$'$, and between
case V and V$'$, respectively.
The functional form of the potential is obtained by substituting
the superpotential (\ref{eqn:sppot}) for Eq.~(\ref{eqn:orham}),
which reads (up to a constant term),
\begin{align}
V(q)
=&\,\frac{1}{2}\sum_{i=1}^{M}\left[\left(
 \frac{\partial\cW}{\partial q_{i}}\right)^{2}-
 \frac{\partial^{2}\cW}{\partial q_{i}^{2}}\right]
 -\bC\bigl(\sigma(h)\bigr)\notag\\
=&\,\sum_{i=1}^{M}\frac{1}{4P(h_{i})}\left[
 \frac{\cN-1+(M-1)c}{2}P'(h_{i})-Q(h_{i})\right]\times\notag\\
&\,\times\left[\frac{\cN+1+(M-1)c}{2}P'(h_{i})-Q(h_{i})\right]
-a_{4}(M,\cN)\sum_{i=1}^{M}h_{i}^{2}-a_{3}(M,\cN)
 \sum_{i=1}^{M}h_{i}\notag\\
&\, +c(c-1)\sum_{i<j}^{M}\left[
 \frac{P(h_{i})+P(h_{j})}{(h_{i}-h_{j})^{2}}-2a_{4}h_{i}h_{j}
 \right],
\label{eqn:potfm}
\end{align}
where the coupling constants $a_{4}(M,\cN)$ and $a_{3}(M,\cN)$
are given by,
\begin{subequations}
\label{eqns:dfa4a3}
\begin{align}
a_{4}(M,\cN)=&\,\bigl[\cN^{2}+2\cN (M-1)c+M(M-1)c^{2}
 -(M-1)c-1\bigr] a_{4},\\
a_{3}(M,\cN)=&\,\bigl[\cN^{2}+2\cN (M-1)c+(M^{2}-1)c^{2}
 -2(M-1)c-1\bigr]\frac{a_{3}}{2}\notag\\
&\, -\bigl[\cN+(M-1)c\bigr]b_{2}.
\end{align}
\end{subequations}
For each case in Table~\ref{tab:class}, we have specific
values of the parameters $a_{p}$ and the transformation function
$h(q)$ obtained by Eq.~(\ref{eqn:solqh}). Substituting all of them
for Eq.~(\ref{eqn:potfm}), we obtain a quasi-solvable potential
in terms of $q_{i}$ in each the case.
Due to the double poles at $h_{i}=h_{j}$ ($i\neq j$)
in the last term of Eq.~(\ref{eqn:potfm}), the two-body potentials
in all the cases have singularities at $q_{i}=q_{j}$ ($i\neq j$).
Hence, each of the models is naturally defined on a Weyl chamber,
\begin{align}
(0<)q_{M}<\dots <q_{1}<\infty,
\end{align}
if the potential is non-periodic on $\bbR^{M}$, or on a Weyl alcove,
\begin{align}
0<q_{M}<\dots <q_{1}<\Omega,
\end{align}
if the potential is periodic on $\bbR^{M}$ with a real period
$\Omega$~\cite{OlPe2}. As we will see below, case I, II, III and
IV${}^{(')}$ with $\nu>0$, and III$'$ with $\nu<0$ correspond to the
former while the other cases, namely, case III and IV${}^{(')}$ with
$\nu<0$, III$'$ with $\nu>0$, and V${}^{(')}$ correspond to the latter.
In the latter case, a system can be quasi-exactly solvable unless
a pole of a one-body potential in the system exists and is in the
Weyl alcove. On the other hand, quasi-exact solvability in the
former case depends mainly on the asymptotic behavior of
Eq.~(\ref{eqn:svwav}) at $|q_{i}|\to\infty$. In both the cases,
finiteness of the contribution from $q_{i}\sim q_{j}$ ($i\neq j$)
to the norm of the wave functions (\ref{eqn:svwav}) requires
$c>-1/2$.

In the following, we will show the functional forms of both
the $\cW(q)$ and $V(q)$ without irrelevant constant terms.

\subsection{\label{ssec:case1}Case I}

In this case, $a_{0}=1/2$ and $a_{1}=a_{2}=a_{3}=a_{4}=0$. From
Eq.~(\ref{eqn:solqh}), we have,
\begin{align}
h(q)=q.
\end{align}
The $\cW(q)$ and $V(q)$ are calculated as follows:
\begin{align}
\cW(q)=-\frac{b_{2}}{3}\sum_{i=1}^{M}q_{i}^{3}-\frac{b_{1}}{2}
 \sum_{i=1}^{M}q_{i}^{2}-b_{0}\sum_{i=1}^{M}q_{i}
 -c\sum_{i<j}^{M}\ln \left|q_{i}-q_{j}\right|,
\label{eqn:sppt1}
\end{align}
\begin{align}
V(q)=&\,\frac{1}{2}\sum_{i=1}^{M}\bigl(b_{2}q_{i}^{2}+b_{1}q_{i}+b_{0}
 \bigr)^{2}+\bigl[\cN+(M-1)c\bigr]b_{2}\sum_{i=1}^{M}q_{i}\notag\\
&\,+c(c-1)\sum_{i<j}^{M}\frac{1}{(q_{i}-q_{j})^{2}}.
\label{eqn:pote1}
\end{align}
This is the rational $A$ type Inozemtsev
model~\cite{Inoze1,InMe1,Inoze3}. Inozemtsev models are known as
a family of deformed CS models which preserve the classical
integrability.
The main difference between quantum and classical case is that the
quantum quasi-solvability holds only for \textit{quantized} values
of the parameters, namely, for integer values of $M$ and $\cN$, while
the classical integrability holds for any continuous values.
This is one of the common features quantum quasi-solvable
models share. The quantum solvability is realized when $b_{2}=0$.
In this case, the model reduces to the rational $A$ type CS model
with the harmonic frequency $b_{1}$. The parameter $b_{0}$ corresponds
to the freedom of the translational invariance and is irrelevant.

\subsection{\label{ssec:case2}Case II}

In this case, $a_{1}=2$ and $a_{0}=a_{2}=a_{3}=a_{4}=0$. From
Eq.~(\ref{eqn:solqh}), we have,
\begin{align}
h(q)=q^{2}.
\end{align}
The $\cW(q)$ and $V(q)$ are calculated as follows:
\begin{align}
\cW(q)=-\frac{b_{2}}{8}\sum_{i=1}^{M}q_{i}^{4}-\frac{b_{1}}{4}
 \sum_{i=1}^{M}q_{i}^{2}-c_{1}\sum_{i=1}^{M}\ln \left|q_{i}\right|
 -c\sum_{i<j}^{M}\ln \left|q_{i}^{2}-q_{j}^{2}\right|,
\label{eqn:sppt2}
\end{align}
\begin{align}
V(q)=&\,\frac{1}{8}\sum_{i=1}^{M}q_{i}^{2}(b_{2}q_{i}^{2}+b_{1})^{2}
 +\frac{1}{4}\bigl[4\cN+4(M-1)c+2c_{1}-1\bigr]b_{2}\sum_{i=1}^{M}
 q_{i}^{2}\notag\\
&\, +\frac{c_{1}(c_{1}-1)}{2}\sum_{i=1}^{M}\frac{1}{q_{i}^{2}}
 +c(c-1)\sum_{i<j}^{M}\left[\frac{1}{(q_{i}-q_{j})^{2}}
 +\frac{1}{(q_{i}+q_{j})^{2}}\right].
\label{eqn:pote2}
\end{align}
The parameter $c_{1}$ is introduced according to,
\begin{align}
2c_{1}=b_{0}-\cN+1-(M-1)c.
\end{align}
This is the rational $BC$ type Inozemtsev model. The quasi-exactly
solvable model reported in Ref.~\cite{HoSh1} is just this case. 
The solvability is realized when $b_{2}=0$.
In this case, the model reduces to the rational $BC$ type CS model
with the harmonic frequency $b_{1}/2$.

\subsection{\label{ssec:case3}Case III}

In this case, $a_{2}=2\nu$ and $a_{0}=a_{1}=a_{3}=a_{4}=0$. From
Eq.~(\ref{eqn:solqh}), we have,
\begin{align}
h(q)=\ee^{2\rnu q}.
\end{align}
The $\cW(q)$ and $V(q)$ are calculated as follows:
\begin{align}
\cW(q)=&\, -\frac{b_{2}}{4\nu}\sum_{i=1}^{M}\ee^{2\rnu q_{i}}
 +\frac{b_{0}}{4\nu}\sum_{i=1}^{M}\ee^{-2\rnu q_{i}}-\rnu
 (\bar{b}_{1}-\cN+1)\sum_{i=1}^{M}q_{i}\notag\\
&\, -c\sum_{i<j}^{M}\ln \left|\sinh\rnu (q_{i}-q_{j})\right|,
\label{eqn:sppt3}
\end{align}
\begin{align}
V(q)=&\,\frac{b_{2}^{2}}{8\nu}\sum_{i=1}^{M}\ee^{4\rnu q_{i}}
 +\frac{b_{2}}{2}\bigl[\bar{b}_{1}+\cN+(M-1)c\bigr]\sum_{i=1}^{M}
 \ee^{2\rnu q_{i}}\notag\\
&\, +\frac{b_{0}^{2}}{8\nu}\sum_{i=1}^{M}\ee^{-4\rnu q_{i}}
 +\frac{b_{0}}{2}\bigl[\bar{b}_{1}-\cN-(M-1)c\bigr]\sum_{i=1}^{M}
 \ee^{-2\rnu q_{i}}\notag\\
&\, +\nu c(c-1)\sum_{i<j}^{M}
 \frac{1}{\sinh^{2}\rnu(q_{i}-q_{j})}.
\label{eqn:pote3}
\end{align}
This is the hyperbolic $(\nu >0)$ and
trigonometric $(\nu <0)$ $A$ type Inozemtsev model. The solvability
is realized when $b_{2}=0$. In this case, the model reduces to
the hyperbolic (trigonometric) $A$ type CS model in the external
Morse potential. The parameter $\bar{b}_{1}=b_{1}/2\nu$ is related
to the translational invariance of the model and is irrelevant.

\subsection{\label{ssec:case3'}Case III$'$}

In this case, $a_{0}=\nu/2$, $a_{2}=\nu$, $a_{4}=\nu/2$ and
$a_{1}=a_{3}=0$. From Eq.~(\ref{eqn:solqh}), we have,
\begin{align}
h(q)=\tan\rnu q.
\end{align}
The $\cW(q)$ and $V(q)$ are calculated as follows:
\begin{align}
\cW(q)=&\, -2\rnu c_{3}\sum_{i=1}^{M}q_{i}-c_{2}\sum_{i=1}^{M}
 \sin 2\rnu q_{i}-c_{1}\sum_{i=1}^{M}\cos 2\rnu q_{i}\notag\\
&\, -(\cN-1)\sum_{i=1}^{M}\ln \left|\cos\rnu q_{i}\right|
 -c\sum_{i<j}^{M}\ln \left|\sin\rnu (q_{i}-q_{j})\right|,
\label{eqn:sppt3'}
\end{align}
\begin{align}
V(q)=&\,\left(c_{2}^{2}-c_{1}^{2}\right)\nu\sum_{i=1}^{M}
 \cos 4\rnu q_{i}+2\nu\Bigl\{2c_{2}c_{3}-\bigl[\cN+(M-1)c\bigr]c_{1}
 \Bigr\}\sum_{i=1}^{M}\cos 2\rnu q_{i}\notag\\
&\, -2c_{1}c_{2}\nu\sum_{i=1}^{M}\sin 4\rnu q_{i}-2\nu\Bigl\{
 2c_{1}c_{3}+\bigl[\cN+(M-1)c\bigr]c_{2}\Bigr\}\sum_{i=1}^{M}
 \sin 2\rnu q_{i}\notag\\
&\, +\nu c(c-1)\sum_{i<j}^{M}\frac{1}{\sin^{2}\rnu (q_{i}-q_{j})}.
\label{eqn:pote3'}
\end{align}
The parameters are introduced according to,
\begin{align}
c_{1}=-\frac{b_{1}}{4\nu},\qquad c_{2}=-\frac{b_{2}-b_{0}}{4\nu},
 \qquad c_{3}=\frac{b_{2}+b_{0}}{4\nu}.
\end{align}
The solvability is realized when $c_{2}=c_{3}$. The parameter
$c_{3}$ is related to the translational invariance of the model
and is irrelevant.
As was mentioned before, this case gives another real function
representation of case III.

\subsection{\label{ssec:case4}Case IV}

In this case, $a_{0}=-2\nu$, $a_{2}=2\nu$ and $a_{1}=a_{3}=a_{4}=0$.
From Eq.~(\ref{eqn:solqh}), we have,
\begin{align}
h(q)=\cosh 2\rnu q.
\end{align}
The $\cW(q)$ and $V(q)$ are calculated as follows:
\begin{align}
\cW(q)=&\, -c_{3}\sum_{i=1}^{M}\cosh 2\rnu q_{i}-c_{2}\sum_{i=1}^{M}
 \ln \left|\sinh 2\rnu q_{i}\right|-c_{1}\sum_{i=1}^{M}\ln \left|
 \sinh\rnu q_{i}\right|\notag\\
&\, -c\sum_{i<j}^{M}\ln \left|\sinh\rnu (q_{i}-q_{j})
 \sinh\rnu (q_{i}+q_{j})\right|,
\label{eqn:sppt4}
\end{align}
\begin{align}
V(q)=&\,2\nu c_{3}^{2}\sum_{i=1}^{M}\sinh^{2}2\rnu q_{i}
 +4\nu c_{3}\bigl[c_{1}+2c_{2}+2(M-1)c\notag\\
&\,+2\cN-1\bigr]\sum_{i=1}^{M}\sinh^{2}\rnu q_{i}
 +2\nu c_{2}(c_{2}-1)\sum_{i=1}^{M}\frac{1}{\sinh^{2}2\rnu q_{i}}\notag\\
&\,+\frac{\nu}{2}c_{1}(c_{1}+2c_{2}-1)\sum_{i=1}^{M}
 \frac{1}{\sinh^{2}\rnu q_{i}}\notag\\
&\,+\nu c(c-1)\sum_{i<j}^{M}\left[\frac{1}{\sinh^{2}\rnu(q_{i}-q_{j})}
 +\frac{1}{\sinh^{2}\rnu(q_{i}+q_{j})}\right].
\label{eqn:pote4}
\end{align}
The parameters $c_{i}$ are introduced according to,
\begin{align}
c_{1}=\frac{b_{0}+b_{2}}{2\nu},\quad c_{2}=\frac{b_{1}-b_{0}-b_{2}}{
 4\nu}-\frac{1}{2}\bigl[\cN-1+(M-1)c\bigr],\quad
 c_{3}=\frac{b_{2}}{4\nu}.
\end{align}
This is the hyperbolic $(\nu >0)$ and
trigonometric $(\nu <0)$ $BC$ type Inozemtsev model.
The solvability is realized when $b_{2}=0$, that is, $c_{3}=0$.
In this case, the model reduces to the hyperbolic (trigonometric)
$BC$ type CS model.
The quasi-solvable models constructed by an ansatz method in
Ref.~\cite{SaTa1} are all included in case I, II, III$'$ and IV.

\subsection{\label{ssec:case4'}Case IV$'$}

In this case, $a_{0}=2\nu$, $a_{2}=2\nu$ and $a_{1}=a_{3}=a_{4}=0$.
From Eq.~(\ref{eqn:solqh}), we have,
\begin{align}
h(q)=\sinh 2\rnu q.
\end{align}
The $\cW(q)$ and $V(q)$ are calculated as follows:
\begin{align}
\cW(q)=&\,-c_{3}\sum_{i=1}^{M}\sinh 2\rnu q_{i}-c_{2}\sum_{i=1}^{M}
 \gd 2\rnu q_{i}-c_{1}\sum_{i=1}^{M}\ln \left|\cosh
 2\rnu q_{i}\right|\notag\\
&\, -c\sum_{i<j}^{M}\ln \left|\sinh\rnu (q_{i}-q_{j})\cosh\rnu
 (q_{i}+q_{j})\right|,
\label{eqn:sppt4'}
\end{align}
\begin{align}
V(q)=&\,2\nu c_{3}^{2}\sum_{i=1}^{M}\cosh^{2}2\rnu q_{i}
 +2\nu c_{3}\bigl[2c_{1}+2(M-1)c+2\cN-1\bigr]\sum_{i=1}^{M}
 \sinh 2\rnu q_{i}\notag\\
&\, +2\nu c_{2}(2c_{1}-1)\sum_{i=1}^{M}
 \frac{\sinh 2\rnu q_{i}}{\cosh^{2}2\rnu q_{i}}+2\nu\bigl[
 c_{2}^{2}-c_{1}(c_{1}-1)\bigr]\sum_{i=1}^{M}
 \frac{1}{\cosh^{2}2\rnu q_{i}}\notag\\
&\, +\nu c(c-1)\sum_{i<j}^{M}\left[
 \frac{1}{\sinh^{2}\rnu (q_{i}-q_{j})}
 -\frac{1}{\cosh^{2}\rnu (q_{i}+q_{j})}\right],
\label{eqn:pote4'}
\end{align}
where $\gd q=\arctan(\sinh q)$ is the Gudermann function.
The parameters $c_{i}$ are introduced according to,
\begin{align}
c_{1}=\frac{b_{1}}{4\nu}-\frac{1}{2}\bigl[\cN-1+(M-1)c\bigr],\quad
 c_{2}=\frac{b_{0}-b_{2}}{4\nu},\quad c_{3}=\frac{b_{2}}{4\nu}.
\end{align}
This form is neither the Inozemtsev nor the Olshanetsky--Perelomov
type~\cite{OlPe2} even in the solvable case, $b_{2}=c_{3}=0$. However,
it can be expressed as the one in case IV if the parameters in
the potential are extended to be complex.
The quasi-solvable models constructed in Ref.~\cite{UlLoRo1} are all
included in case I, II, III, IV and IV$'$.

\subsection{\label{ssec:case5}Case V}

In this case, $a_{0}=-g_{3}/2$, $a_{1}=-g_{2}/2$, $a_{3}=2$ and
$a_{2}=a_{4}=0$. From Eq.~(\ref{eqn:solqh}), we have,
\begin{align}
h(q)=\wp (q;g_{2},g_{3}),
\end{align}
where $\wp(q;g_{2},g_{3})$ is the Weierstrass elliptic function with
the invariants $g_{2}$ and $g_{3}$, and will be abbreviated to $\wp(q)$
hereafter. The $\cW(q)$ and $V(q)$ are calculated as follows:
\begin{align}
\cW(q)
=&\, -b_{1}\sum_{i=1}^{M}\int\dd q_{i}\frac{\wp(q_{i})}{\wp'(q_{i})}
 -\bar{b}_{0}\sum_{i=1}^{M}\int\frac{\dd q_{i}}{\wp'(q_{i})}
 -c_{1}\sum_{i=1}^{M}\ln \left|\wp'(q_{i})\right|\notag\\
&\, -c\sum_{i<j}^{M}\ln \left|\wp(q_{i})-\wp(q_{j})\right|.
\label{eqn:sppt51}
\end{align}
\begin{align}
V(q)=&\,\alpha_{1}\sum_{i=1}^{M}\frac{\wp''(q_{i})}{\wp'(q_{i})^{2}}
 +\alpha_{2}\sum_{i=1}^{M}\frac{\wp(q_{i})}{\wp'(q_{i})^{2}}
 +\alpha_{3}\sum_{i=1}^{M}\frac{1}{\wp'(q_{i})^{2}}
 +c_{2}(M,\cN)\sum_{i=1}^{M}\wp(q_{i})\notag\\
&\, +2c_{1}(c_{1}-1)\sum_{i=1}^{M}\wp(2q_{i})+c(c-1)\sum_{i<j}^{M}
 \bigl[\wp(q_{i}-q_{j})+\wp(q_{i}+q_{j})\bigr].
\label{eqn:pote51}
\end{align}
The parameters $\bar{b}_{0}$, $c_{1}$ are given by,
\begin{align}
\bar{b}_{0}=b_{0}+\frac{b_{2}g_{2}}{12},\qquad
 c_{1}=\frac{b_{2}}{6}-\frac{1}{2}\bigl[\cN-1+(M-1)c\bigr],
\end{align}
and the coupling constants $\alpha_{i}$ and $c_{2}(M,\cN)$ are
defined by,
\begin{subequations}
\begin{align}
\alpha_{1}&=\frac{1}{12}\bigl[6\bar{b}_{0}(2c_{1}-1)+b_{1}^{2}\bigr],\\
\alpha_{2}&=\frac{b_{1}}{2}\bigl[(2c_{1}-1)g_{2}+2\bar{b}_{0}\bigr],\\
\alpha_{3}&=\frac{1}{24}\bigl[18b_{1}(2c_{1}-1)g_{3}+12\bar{b}_{0}^{2}
 +b_{1}^{2}g_{2}\bigr],
\end{align}
\end{subequations}
\begin{align}
c_{2}(M,\cN)&=\bigl[\cN-1+(M-1)c+2c_{1}\bigr]\bigl[2\cN-1+2(M-1)c
 +2c_{1}\bigr].
\end{align}
The above expressions (\ref{eqn:sppt51})--(\ref{eqn:pote51}) can
be arranged into a more familiar form if we introduce the values
of $\wp(q)$ at the half of the fundamental periods $2\omega_{l}$:
\begin{align}
e_{l}=\wp(\omega_{l})\qquad (l=1,2,3).
\label{eqn:vlhfp}
\end{align}
Utilizing the well-known formulae of the Weierstrass function:
\begin{subequations}
\label{eqns:fmwp1}
\begin{align}
2P(h)&=\wp'(q)^{2}=4\bigl(\wp(q)-e_{1}\bigr)
 \bigl(\wp(q)-e_{2}\bigr)\bigl(\wp(q)-e_{3}\bigr),\\
2P'(h)&=2\wp''(q)=12\wp(q)^{2}-g_{2},
\end{align}
\end{subequations}
and the addition theorem,
\begin{align}
\wp(q_{i}+\omega_{l})=\frac{H_{l}^{2}}{\wp(q_{i})-e_{l}}+e_{l},
\label{eqn:addit}
\end{align}
we have the following forms of $\cW(q)$ and $V(q)$:
\begin{align}
\cW(q)=&\,\sum_{l=1}^{3}\frac{b_{1}e_{l}-\bar{b}_{0}}{4H_{l}^{2}}
 \sum_{i=1}^{M}\ln\left|\wp(q_{i})-e_{l}\right|
 -c_{1}\sum_{i=1}^{M}\ln\left|\wp'(q_{i})\right|\notag\\
&\, -c\sum_{i<j}^{M}\ln \left|\wp(q_{i})-\wp(q_{j})\right|.
\label{eqn:sppt52}
\end{align}
\begin{align}
V(q)=&\,\sum_{l=1}^{3}\frac{2\alpha_{1}H_{l}^{2}-\alpha_{2}
 e_{l}+\alpha_{3}}{4H_{l}^{4}}\sum_{i=1}^{M}\wp(q_{i}+\omega_{l})
 +c_{2}(M,\cN)\sum_{i=1}^{M}\wp(q_{i})\notag\\
&\, +2c_{1}(c_{1}-1)\sum_{i=1}^{M}\wp(2q_{i})+c(c-1)\sum_{i<j}^{M}
 \bigl[\wp(q_{i}-q_{j})+\wp(q_{i}+q_{j})\bigr].
\label{eqn:pote52}
\end{align}
In the above, $H_{l}$ are defined by,
\begin{align}
H_{l}^{2}=(e_{l}-e_{m})(e_{l}-e_{n})=3e_{l}^{2}-\frac{g_{2}}{4}
 \qquad(l=1,2,3;\ l\neq m\neq n\neq l).
\end{align}
The potential form (\ref{eqn:pote52}) shows that this case is
the elliptic $BC$ type Inozemtsev model.
Contrary to all the previous cases I--IV$'$, the solvability is not
realized for any values of the free parameters because $a_{3}\neq 0$.
On the other hand, the model becomes the elliptic $BC$ type CS model
when $\alpha_{1}=\alpha_{2}=\alpha_{3}=0$, or equivalently,
$\bar{b}_{0}=b_{1}=0$. Therefore, in contrast to the fact that
the rational and hyperbolic (trigonometric) $A$ and $BC$ type CS models
are solvable as quantum systems, the elliptic $BC$ type CS model can be
only quasi-solvable but is not solvable as a quantum system.
The quantum elliptic model investigated in Ref.~\cite{UlLoRo2,BrHa1}
is just the CS model with the translation $q_{i}\to q_{i}+i\beta$.
We note that from the formula,
\begin{align}
\wp(q_{i}+w_{l})=\wp_{l}^{(1/2)}(q_{i})-\wp(q_{i})+e_{l},
\end{align}
where $\wp_{l}^{(1/2)}(q)$ denotes the Weierstrass function with one
of the fundamental periods $2\omega_{l}$ replaced by $\omega_{l}$,
the above model is related to the twisted elliptic CS
models~\cite{HoPh1,HoPh2,BoSa1}.

\subsection{\label{ssec:case5'}Case V$'$}

In this case, $a_{0}=\nu/2$, $a_{2}=\nu(2-k^{2})/2$,
$a_{4}=\nu(1-k^{2})/2$ and $a_{1}=a_{3}=0$. From Eq.~(\ref{eqn:solqh}),
we have,
\begin{align}
h(q)=\frac{\sn(\rnu q|k)}{\cn(\rnu q|k)},
\end{align}
where $\sn(\rnu q|k)$ etc. are the Jacobian elliptic functions with
the modulus $k$, and will be abbreviated to $\sn(\rnu q)$ etc.
hereafter. The $\cW(q)$ and $V(q)$ are calculated as follows:
\begin{align}
\cW(q)=&\, -\frac{b_{2}}{\rnu}\sum_{i=1}^{M}\int\dd q_{i}
 \frac{\mathrm{sn}^{2}\,\rnu q_{i}}{\dn\rnu q_{i}}-\frac{b_{0}}{\rnu}
 \sum_{i=1}^{M}\int\dd q_{i}\frac{\mathrm{cn}^{2}\,\rnu q_{i}}{
 \dn\rnu q_{i}}\notag\\
&\, +\left(\frac{b_{1}}{k^{2}\nu}+\frac{\cN-1+c_{M}}{2}\right)
 \sum_{i=1}^{M}\ln\left|\dn\rnu q_{i}\right|-(\cN-1)\sum_{i=1}^{M}
 \ln\left|\cn\rnu q_{i}\right|\notag\\
&\, -c\sum_{i<j}^{M}\ln\left|\sn\rnu q_{i}\,\cn\rnu q_{j}
 -\cn\rnu q_{i}\,\sn\rnu q_{j}\right|,
\label{eqn:sppt5'}
\end{align}
\begin{align}
V(q)=&\,\alpha_{1}\sum_{i=1}^{M}\sn\rnu q_{i}\cn\rnu q_{i}+\alpha_{2}
 \sum_{i=1}^{M}\sn^{2}\rnu q_{i}+\sum_{i=1}^{M}
 \frac{\alpha_{3}\sn\rnu q_{i}\cn\rnu q_{i}+\alpha_{4}}{\dn^{2}\rnu q_{i}}
 \notag\\
&\,+\frac{\nu c(c-1)}{2}\sum_{i<j}^{M}\frac{\dn^{2}\rnu q_{i}+\dn^{2}
 \rnu q_{j}}{(\sn\rnu q_{i}\cn\rnu q_{j}-\cn\rnu q_{i}\sn\rnu q_{j})^{2}},
\label{eqn:pote5'}
\end{align}
where $c_{M}=(M-1)c$ and the coupling constants $\alpha_{i}$ are given
by,
\begin{align}
\alpha_{1}=&\,\frac{b_{0}-b_{2}}{2k^{2}\nu}\bigl[2b_{1}-(\cN+c_{M})k^{2}
 \nu\bigr],\\
\alpha_{2}=&\,\frac{1}{2k^{2}\nu}\Biggl[\biggl(b_{1}-\frac{\cN-1+c_{M}}{2}
 k^{2}\nu\biggr)\biggl(b_{1}-\frac{\cN+1+c_{M}}{2}k^{2}\nu\biggr)
-(b_{0}-b_{2})^{2}\Biggr],\\
\alpha_{3}=&\,-\frac{k^{\prime 2}b_{0}-b_{2}}{2k^{2}\nu}\bigl[2b_{1}
 +(\cN+c_{M})k^{2}\nu\bigr],\\
\alpha_{4}=&\,-\frac{1}{2k^{4}\nu}\Biggl[k^{\prime 2}\biggl(b_{1}
 +\frac{\cN-1+c_{M}}{2}k^{2}\nu\biggr)\biggl(b_{1}+\frac{\cN+1+c_{M}}{2}
 k^{2}\nu\biggr)
-(k^{\prime 2}b_{0}-b_{2})^{2}\Biggr],
\end{align}
where $k^{\prime 2}=1-k^{2}$.
The solvability is not realized for any values of the free parameters
because $a_{4}\neq 0$.
As was mentioned before, this case gives another real function
representation of case V.

\section{\label{sec:possi}Possibility Beyond Two-body Interactions}

So far, we have only concerned with up to two-body interactions.
In this section, we will investigate what kind of many-body
interactions can be constructed from our procedure. It can be
achieved, in principle, by solving the canonical-form condition
(\ref{eqn:cfcon}) for $M=3,4,5,\ldots$. However, we will see that
the $GL(2,K)$ symmetry provide a powerful tool for this problem.

Let us first consider the $M=3$ case. From the same procedure as in
Eqs.~(\ref{eqn:alge2})--(\ref{eqn:chai2}), we obtain the transforms
of the differential operators:
\begin{align}
\frac{\partial}{\partial\sigma_{l}}=\sum_{i=1}^{3}
 \frac{(-h_{i})^{3-l}}{(h_{i}-h_{j})(h_{i}-h_{k})}
 \frac{\partial}{\partial h_{i}}\qquad(l=1,2,3;\ i\neq j\neq k\neq i).
\label{eqn:chai3}
\end{align}
As we have seen in Section~\ref{sec:canon}, the second-order
differential operators in the gauged Hamiltonian must be one-body
in order to satisfy the canonical-form condition. So, three-body
operators can only exist as first- and zeroth-order differential
operators. From the form of the gauged Hamiltonian Eq.~(\ref{eqn:sgham})
for $M=3$ with Eq.~(\ref{eqn:chai3}), we can conclude that only the
following form of the three-body operator may be added as first-order
differential operators to the $M$-body gauged Hamiltonian
(\ref{eqn:htham}):
\begin{align}
\sum_{i\neq j\neq k\neq i}^{M}\frac{O(h_{i})}{(h_{i}-h_{j})(h_{i}-h_{k})}
 \frac{\partial}{\partial h_{i}},
\label{eqn:3bdop}
\end{align}
where $O(h_{i})$ denotes a polynomial. On the other hand,
$\tH_{\cN}$ as a whole should be shape invariant under the
$GL(2,K)$ transformation (\ref{eqn:tfham}). From the formulae
(\ref{eqn:ivdtf}) and (\ref{eqn:1dvtf}), the operator (\ref{eqn:3bdop})
is transformed as,
\begin{align}
\text{Eq.~(\ref{eqn:3bdop})}
\mapsto&\,
\sum_{i\neq j\neq k\neq i}^{M}\frac{\Delta^{-3}(\gamma h_{i}
 +\delta)^{6}O(\hat{h}_{i})}{(h_{i}-h_{j})(h_{i}-h_{k})}
 \frac{\partial}{\partial h_{i}}-2(M-2)\sum_{i\neq j}^{M}
 \frac{\Delta^{-3}\gamma(\gamma h_{i}+\delta)^{5}O(\hat{h}_{i})}{
 h_{i}-h_{j}}\frac{\partial}{\partial h_{i}}\notag\\
&\, +(M-1)(M-2)\sum_{i=1}^{M}
 \Delta^{-3}\gamma^{2}(\gamma h_{i}+\delta)^{4}O(\hat{h}_{i})
 \frac{\partial}{\partial h_{i}}+\text{(0th-order ops.)},
\label{eqn:tf3op}
\end{align}
where (0th-order ops.) denotes the terms which do not involve any
differential operators.
From the first term of the r.h.s. in Eq.~(\ref{eqn:tf3op}), the
requirement of the $GL(2,K)$ invariance determines the transformation
of the $O(h_{i})$:
\begin{align}
O(h_{i})\mapsto \hat{O}(h_{i})=\Delta^{-3}(\gamma h_{i}+\delta)^{6}
 O(\hat{h}_{i}).
\label{eqn:tfofO}
\end{align}
This means that $O(h)$ should be a polynomial of sixth-degree.
To complete the invariance, the second and third terms of the r.h.s.
in Eq.~(\ref{eqn:tf3op}) should be absorbed into the other two-body
and one-body operators, respectively. We find that these operators
should be of the following forms:
\begin{align}
\sum_{i\neq j}^{M}\frac{O'(h_{i})}{h_{i}-h_{j}}
 \frac{\partial}{\partial h_{i}},\qquad
\sum_{i=1}^{M}O''(h_{i})\frac{\partial}{\partial h_{i}},
\label{eqn:muexo}
\end{align}
since the derivatives of $O(h_{i})$ transform according to
Eq.~(\ref{eqns:dvofO}).
Similar observation tells us what kind of zeroth-order operators should
simultaneously exist. Finally, we find the $GL(2,K)$ invariant
combination of the three-body operators:
\begin{align}
\tH_{\cN}^{(3)}=&\,
\sum_{i\neq j\neq k\neq i}^{M}\frac{O(h_{i})}{(h_{i}-h_{j})(h_{i}-h_{k})}
 \frac{\partial}{\partial h_{i}}-\frac{M-2}{3}\sum_{i\neq j}^{M}
 \frac{O'(h_{i})}{h_{i}-h_{j}}\frac{\partial}{\partial h_{i}}
\notag\\
&\, +\frac{(M-1)(M-2)}{30}\sum_{i=1}^{M}O''(h_{i})\frac{\partial}{
 \partial h_{i}}-(\cN-1)\left[\frac{1}{6}\sum_{i\neq j\neq k\neq i}^{M}
 \frac{O'(h_{i})}{(h_{i}-h_{j})(h_{i}-h_{k})}\right.\notag\\
&\, -\left.\frac{M-2}{15}\sum_{i\neq j}^{M}\frac{O''(h_{i})}{h_{i}-h_{j}}
 +\frac{(M-1)(M-2)}{120}\sum_{i=1}^{M}O'''(h_{i})\right].
\label{eqn:inv3o}
\end{align}
Therefore, the three-body terms should be added to the gauged Hamiltonian
(\ref{eqn:htham}) as the constant multiple of the combination
(\ref{eqn:inv3o}). The form of the total gauged Hamiltonian reads,
\begin{align}
\tH_{\cN}+g_{3}\tH_{\cN}^{(3)}=-\sum_{i=1}^{M}P(h_{i})
 \frac{\partial^{2}}{\partial h_{i}^{2}}+\sum_{i=1}^{M}S_{i}(h)
 \frac{\partial}{\partial h_{i}}-T(h),
\label{eqn:totgH}
\end{align}
where $S_{i}(h)$ is given by,
\begin{align}
\lefteqn{
S_{i}(h)=\frac{\cN-2+(M-1)c}{2}P'(h_{i})-Q(h_{i})-2c\sum_{j(\neq i)}^{M}
 \frac{P(h_{i})}{h_{i}-h_{j}}
}\notag\\
& +g_{3}\left[\sum_{(i\neq)j\neq k(\neq i)}^{M}
 \frac{O(h_{i})}{(h_{i}-h_{j})(h_{i}-h_{k})}
 -\frac{M-2}{3}\sum_{j(\neq i)}^{M}\frac{O'(h_{i})}{h_{i}-h_{j}}
 +\frac{(M-1)(M-2)}{30}O''(h_{i})\right].
\label{eqn:defSi}
\end{align}
In order that the total gauged Hamiltonian (\ref{eqn:totgH}) can be
cast in the Schr\"{o}dinger form by a gauge transformation, we must have,
\begin{align}
\frac{\partial\cW}{\partial q_{i}}=\frac{h''_{i}}{2h'_{i}}
 +\frac{S_{i}(h)}{h'_{i}}\qquad (i=1,\dots,M),
\end{align}
in addition to Eq.~(\ref{eqn:dhofq}). Thus, the integrability condition
for $\cW(q)$ becomes,
\begin{align}
\frac{\partial}{\partial q_{l}}\frac{\partial\cW}{\partial q_{i}}
=\frac{\partial}{\partial q_{i}}\frac{\partial\cW}{\partial q_{l}}
\quad\Leftrightarrow\quad
\frac{h'_{l}}{h'_{i}}\frac{\partial S_{i}(h)}{\partial h_{l}}
 =\frac{h'_{i}}{h'_{l}}\frac{\partial S_{l}(h)}{\partial h_{i}}
 \qquad (\forall\,i\neq l).
\end{align}
However, this condition cannot be satisfied unless $g_{3}=0$ since
we obtain from Eq.~(\ref{eqn:defSi}),
\begin{align}
\frac{h'_{l}}{h'_{i}}\frac{\partial S_{i}(h)}{\partial h_{l}}
 =-c\frac{h'_{i}h'_{l}}{(h_{i}-h_{l})^{2}}
 +\frac{g_{3}}{(h_{i}-h_{l})^{2}}
 \frac{h'_{l}}{h'_{i}}\left[\sum_{j(\neq i)}^{M}
 \frac{2O(h_{i})}{h_{i}-h_{j}}-\frac{M-2}{3}O'(h_{i})\right].
\end{align}
In other words, the $\fsl(M+1)$ quasi-solvable operator (\ref{eqn:sgham})
which satisfies the canonical-form condition cannot contain the
three-body operators.

The generalization of the above argument to the $M\ge 3$ cases
is straightforward and results in the same conclusion that the
existence of $M$-body operators for $M\ge 3$ is prohibited. Therefore,
the quasi-solvable quantum many-body systems constructed from the
procedure in Section~\ref{sec:algeb} are exhausted by the ones
classified in the preceding section.

\section{\label{sec:concl}Concluding Remarks}

In conclusion, we have proposed a systematic method to construct
quasi-solvable quantum many-body systems. In Table~\ref{tab:model},
we summarize the complete list of the quasi-solvable models as well
as their special subclass of the solvable models constructed in
this paper.

\begin{table}[h]
\begin{center}
\tabcolsep=5mm
\begin{tabular}{|c|c|c|}\hline
Case & Quasi-solvable & Solvable \\
\hline
I & rational $A$ Inozemtsev & rational $A$ Calogero--Sutherland \\
\hline
II & rational $BC$ Inozemtsev & rational $BC$ Calogero--Sutherland \\
\hline
III & hyp. (trig.) $A$ Inozemtsev & hyp. (trig.) $A$ Calogero--Sutherland \\
 & & + external Morse potential \\
\hline
IV &  hyp. (trig.) $BC$ Inozemtsev & hyp. (trig.) $BC$ Calogero--Sutherland \\
\hline
V & elliptic $BC$ Inozemtsev & {}\hrulefill \\
  & (elliptic $BC$ Calogero--Sutherland) & \\
\hline
\end{tabular}
\caption{Classification of the (quasi-)solvable quantum many-body
 systems.}
\label{tab:model}
\end{center}
\end{table}

We would like to close this paper by giving several remarks
on the future problems.

\begin{enumerate}

\item We should stress that quasi-solvable quantum many-body models
 which can be constructed from $\fsl(M+1)$ generators are not
 limited to the ones presented here. This is because it depends on
 the way of changing of the variables and on the choice of the
 solvable subspace $\cV_{\cN}$. In Ref.~\cite{MiRoTu1}, several
 $M$-body quasi-solvable quantum models with up to $M$- and up to
 6-body interactions were constructed from $\fsl(M+1)$ generators.
 The Calogero--Marchioro--Wolfes model~\cite{Wolfe1,CaMa1} or the
 rational $G_{2}$ type CS model and the trigonometric $G_{2}$ type CS
 model without the two-body interaction, all of which have
 three-body interactions, were shown to have $\fsl(3)$
 algebraizations and thus to be solvable in Refs.~\cite{Quesne1,%
CaRoTu1,Turbi4}. Therefore, there may be another family of
 quasi-solvable quantum many-body systems which can be constructed
 from a different $\fsl(M+1)$ algebraization.

\item From Table~\ref{tab:model}, we notice that, among the classical
 type CS models, only the elliptic $A$ type CS model cannot be obtained
 from the present procedure. As far as we know, the quantum
 (quasi-)solvability of this model has not been confirmed yet.
 We note that this model is also a special in the sense that any
 deformation of the model preserving the integrability have
 not been discovered yet in contrast to the other classical type
 CS models.

\item Table~\ref{tab:model} indicates that classical integrability may
 have more intimate relation with quantum quasi-solvability rather than
 with quantum solvability and quasi-exact solvability. The latter
 is understood because classical integrability does not depend on
 whether the particles are moving inside a bounded region or not.
 So, it will be an interesting problem to study the relation between
 classical integrability and quantum quasi-solvability.

\item Although the paper only deals with scalar problems without
 any internal degree of freedom, there have been much progress
 in constructing quasi-solvable spin systems recently. Among them,
 the generalized Dunkl operator approaches have provided one of
 the most powerful tools for the issues~\cite{FULRZ1,FULRZ2,FULRZ3}.
 Then, it is quite interesting to note that the spin CS models
 constructed from the $A_{M}$-type Dunkl operators~\cite{FULRZ1}
 coincide, in the scalar limit, with the (scalar) models constructed
 from the $\fsl(M+1)$ generators in this paper. Thus, there should
 be some intimate relations underlying between the two approaches
 although the algebraic structure of the $A_{M}$-type Dunkl operators
 and that of the $\fsl(M+1)$ are quite different from each other.
 Another interesting problem is the possibility to generalize
 the Dunkl operators to include $M$-body ($M\ge 3$) operators.
 The conclusion in Section~\ref{sec:possi} however indicates
 little possibility for such a generalization, at least, of the
 $A_{M}$-type.

\item In this paper, we imposed the canonical-form condition
 (\ref{eqn:cfcon}) in order to obtain the ordinary Schr\"{o}dinger
 operators. However, it may be possible to obtain another class
 of quasi-solvable second-order operators by imposing another type
 of condition instead of Eq.~(\ref{eqn:cfcon}).
 If we can solve the canonical-form condition on a specific
 non-trivial metric or on a particular manifold, we will obtain
 a gravitational deformation of the quasi-solvable quantum models.

\item A generalization of the Bender--Dunne polynomials to the
 models obtained here will be an interesting problem. Although It was
 argued in Ref.~\cite{KrUsWa1} that they exist for any quasi-solvable
 quantum systems regardless of the number of particles, we have not
 appreciated whether their arguments can be naively applicable to our
 case.

\item As we have discussed in the end of Section~\ref{sec:gl2ks},
 all the solvable models constructed in this paper are supersymmetric.
 It was shown in Refs.~\cite{BrTuWy1,FrMe1,Shast1} that the rational
 and hyperbolic (trigonometric) $A$ and $BC$ type CS models are of the
 supersymmetric form. Table~\ref{tab:model} shows that the
 hyperbolic (trigonometric) $A$ type CS model can be deformed without
 destroying both the solvability and supersymmetry by adding the
 external Morse potential. Then, a natural question arises whether
 there is a solvable quantum model which is not of the supersymmetric
 form.

\item In the case of $M=1$, it was shown that quasi-solvability and
 $\cN$-fold supersymmetry is essentially equivalent~\cite{AST2}.
 However, the relation between them has not been understood in
 many-body quantum systems although $\cN$-fold supersymmetry in
 many-body case was briefly formulated in Ref.~\cite{AST2}.
 We expect that they also have the similar intimate relation to
 each other in the many-body case.

\end{enumerate}

\begin{acknowledgments}
The author would like to thank H.~Aoyama, F.~Finkel,
D.~G\'{o}mez-Ullate, A.~Gonz\'{a}lez-L\'{o}pez, N.~Nakayama,
M.~A.~Rodr\'{\i}guez, R.~Sasaki, M.~Sato, K.~Takasaki and
S.~Yamaguchi for useful discussions and thank Y.~Hosotani for careful
reading of the manuscript.
The author would also like to thank all the members of Departamento
de F\'{\i}sica Te\'{o}rica II, Universidad Complutense, for their
kind hospitality during his stay.
This work was supported in part by a JSPS research fellowship.
\end{acknowledgments}

\appendix

\section{\label{sec:fmlGL}Formulae for $GL(2,K)$ Transformation}

Here we will list the formulae useful for the calculation of
$GL(2,K)$ transformation in Section~\ref{sec:gl2ks}. Under the
linear fractional transformation (\ref{eqn:frtsf}), we have,
\begin{gather}
h_{i}-h_{j}\mapsto \hat{h}_{i}-\hat{h}_{j}=
 \frac{\Delta (h_{i}-h_{j})}{(\gamma h_{i}+\delta)
 (\gamma h_{j}+\delta)},
\label{eqn:diftf}\\
\frac{1}{h_{i}-h_{j}}\mapsto\frac{1}{\hat{h}_{i}-\hat{h}_{j}}
 =\Delta^{-1}(\gamma h_{i}+\delta)^{2}\left[\frac{1}{h_{i}-h_{j}}
 -\gamma (\gamma h_{i}+\delta)^{-1}\right].
\label{eqn:ivdtf}
\end{gather}
The differential operators in the gauged Hamiltonian are transformed
according to,
\begin{align}
\frac{\partial}{\partial h_{i}}
\mapsto&\,\prod_{j=1}^{M}(\gamma h_{j}+\delta)^{\cN-1}
 \frac{\partial}{\partial \hat{h}_{i}}\prod_{j=1}^{M}
 (\gamma h_{j}+\delta)^{-(\cN-1)}\notag\\
&\, =\Delta^{-1}(\gamma h_{i}+\delta)^{2}\left[
 \frac{\partial}{\partial h_{i}}-(\cN-1)\gamma(\gamma h_{i}
 +\delta)^{-1}\right],
\label{eqn:1dvtf}
\end{align}
and,
\begin{align}
\frac{\partial^{2}}{\partial h_{i}^{2}}
\mapsto&\,\prod_{j=1}^{M}(\gamma h_{j}+\delta)^{\cN-1}
 \frac{\partial^{2}}{\partial \hat{h}_{i}^{2}}\prod_{j=1}^{M}
 (\gamma h_{j}+\delta)^{-(\cN-1)}\notag\\
&\, =\Delta^{-2}(\gamma h_{i}+\delta)^{4}\biggl[
 \frac{\partial^{2}}{\partial h_{i}^{2}}-2(\cN-2)\gamma
 (\gamma h_{i}+\delta)^{-1}\frac{\partial}{\partial h_{i}}\notag\\
&\, \quad +(\cN-1)(\cN-2)\gamma^{2}
 (\gamma h_{i}+\delta)^{-2}\biggr].
\label{eqn:2dvtf}
\end{align}
From the transformation rule (\ref{eqns:tfpol}) for the polynomials
$P(h)$ and $Q(h)$, the derivatives of the transformed polynomials
are calculated as,
\begin{subequations}
\label{eqns:dvptfs}
\begin{align}
\hat{P}'(h_{i})
&=\Delta^{-1}(\gamma h_{i}+\delta)^{2}P'(\hat{h}_{i})
 +4\Delta^{-2}\gamma (\gamma h_{i}+\delta)^{3}P(\hat{h}_{i}),\\
\hat{P}''(h_{i})
&=P''(\hat{h}_{i})+6\Delta^{-1}\gamma (\gamma h_{i}+\delta)
 P'(\hat{h}_{i})+12\Delta^{-2}\gamma^{2}(\gamma h_{i}+\delta)^{2}
 P(\hat{h}_{i}),\\
\hat{Q}'(h_{i})&=Q'(\hat{h}_{i})+2\Delta^{-1}\gamma
 (\gamma h_{i}+\delta)Q(\hat{h}_{i}).
\end{align}
\end{subequations}
Similarly, the derivatives of the transformed sextic polynomial
$\hat{O}(h)$ given by Eq.~(\ref{eqn:tfofO}) are calculated as,
\begin{subequations}
\label{eqns:dvofO}
\begin{align}
\hat{O}'(h_{i})
=&\,\Delta^{-2}(\gamma h_{i}+\delta)^{4}O'(\hat{h}_{i})
 +6\Delta^{-3}\gamma(\gamma h_{i}+\delta)^{5}O(\hat{h}_{i}),\\
\hat{O}''(h_{i})
=&\,\Delta^{-1}(\gamma h_{i}+\delta)^{2}O''(\hat{h}_{i})
 +10\Delta^{-2}\gamma(\gamma h_{i}+\delta)^{3}O'(\hat{h}_{i})\notag\\
&\, +30\Delta^{-3}\gamma^{2}(\gamma h_{i}+\delta)^{4}
 O(\hat{h}_{i}),\\
\hat{O}'''(h_{i})
=&\, O'''(\hat{h}_{i})+12\Delta^{-1}\gamma(\gamma h_{i}+\delta)
 O''(\hat{h}_{i})\notag\\
&\, +60\Delta^{-2}\gamma^{2}(\gamma h_{i}+\delta)^{2}
 O'(\hat{h}_{i})+120\Delta^{-3}\gamma^{3}(\gamma h_{i}+\delta)^{3}
 O(\hat{h}_{i}).
\end{align}
\end{subequations}

\section{\label{sec:tflIP}Transformation Formulae between $h_{i}$
 and $\sigma_{i}$}

In this appendix, we will show a systematic method to calculate the
transformation formulae between the variables $h_{i}$ and $\sigma_{i}$
needed in Section~\ref{sec:inver} and \ref{sec:possi}. The method
presented here is essentially based on the one in Ref.~\cite{MiRoTu1}.
The starting point is that, for the elementary symmetric polynomials
$\sigma_{k}$ and the Newton polynomials $s_{n}$ defined by,
\begin{align}
\sigma_{k}(h)=\sum_{i_{1}<\dots <i_{k}}h_{i_{1}}\cdots h_{i_{k}},
 \quad\sigma_{0}=1;\qquad s_{n}(h)=\sum_{i=1}^{M}h_{i}^{n},
\label{eqn:symvs}
\end{align}
the following relation holds:
\begin{align}
G(t)=\sum_{k=0}^{M}\sigma_{k}t^{k}=\exp\left(\sum_{n=1}^{\infty}
 \frac{(-)^{n+1}}{n}s_{n}t^{n}\right).
\label{eqn:geneG}
\end{align}
Thus, $G(t)$ gives the generating function for both the elementary
symmetric polynomials and the Newton ones. The derivatives of $G(t)$
with respect to $h_{i}$ and $t$ are,
\begin{subequations}
\label{eqns:formuG}
\begin{align}
\frac{\partial}{\partial h_{i}}G(t)
&=G(t)\sum_{n=1}^{\infty}(-)^{n+1}h_{i}^{n-1}t^{n},
\label{eqn:formG1}\\
\frac{\partial}{\partial t}G(t)
&=G(t)\sum_{n=1}^{\infty}(-)^{n+1}t^{n-1}s_{n}
 =\sum_{k=0}^{M}k\sigma_{k}t^{k-1},
\label{eqn:formG2}\\
\frac{\partial^{2}}{\partial t^{2}}G(t)
&=G(t)\left(\sum_{m,m'=1}^{\infty}(-)^{m+m'+2}t^{m+m'-2}s_{m}s_{m'}
 -\sum_{m=1}^{\infty}(-)^{m+1}t^{m-1}m\, s_{m+1}\right).
\label{eqn:formG3}
\end{align}
\end{subequations}
We will see that most of the formulae involving differential operators
can be derived with the aid of the generating function $G(t)$.
Here we derive a useful formula for the later purposes. Let us define
a generating function $B(t)$ as,
\begin{align}
B(t)=\sum_{k=0}^{M}\frac{\partial\sigma_{k}}{\partial h_{i}}t^{k}.
\label{eqn:form21}
\end{align}
From Eq.~(\ref{eqn:formG1}), we have another expression for $B(t)$:
\begin{align}
B(t)&=\frac{\partial}{\partial h_{i}}G(t)
=t\, G(t)-\sum_{n=2}^{\infty}(-)^{n}h_{i}^{n-1}t^{n}G(t)
\notag\\
&=\left(1-h_{i}\frac{\partial}{\partial h_{i}}\right)t\, G(t)
 =\left(1-h_{i}\frac{\partial}{\partial h_{i}}\right)
 \sum_{k=0}^{M} \sigma_{k-1}t^{k}.
\label{eqn:form22}
\end{align}
Comparing Eq.~(\ref{eqn:form21}) and the last expression in
Eq.~(\ref{eqn:form22}), we have the following formula,
\begin{align}
\frac{\partial\sigma_{k}}{\partial h_{i}}=\sigma_{k-1}
 -h_{i}\frac{\partial\sigma_{k-1}}{\partial h_{i}},
\label{eqn:form23}
\end{align}
which facilitates the derivation of the several formulae.

\subsection{\label{ssec:B-p}Calculation of $B_{-}^{(p)}(\sigma)$}

The definition of the quantity $B_{-}^{(p)}$ is given by,
\begin{align}
\sum_{i=1}^{M}h_{i}^{p}=s_{p}(h)=B_{-}^{(p)}(\sigma).
\label{eqn:for3-0}
\end{align}
From the formula (\ref{eqn:formG2}), we can immediately obtain,
\begin{align}
B_{-}^{(p)}
=\frac{(-)^{p-1}}{(p-1)!}\frac{\partial^{p-1}}{\partial t^{p-1}}
 \left[G(t)^{-1}\frac{\partial}{\partial t}G(t)\right]_{t=0}.
\label{eqn:for3-1}
\end{align}
The first few formulae read,
\begin{subequations}
\label{eqns:formB-}
\begin{align}
B_{-}^{(0)}&=s_{0}=M,\\
B_{-}^{(1)}&=s_{1}=\sigma_{1},\\
B_{-}^{(2)}&=s_{2}=\sigma_{1}^{2}-2\sigma_{2},\\
B_{-}^{(3)}&=s_{3}=\sigma_{1}^{3}-3\sigma_{2}\sigma_{1}+3\sigma_{3}.
\end{align}
\end{subequations}

\subsection{\label{ssec:Bkp}Calculation of $B_{k}^{(p)}(\sigma)$}

The definition of the quantity $B_{k}^{(p)}$ is given by,
\begin{align}
\sum_{i=1}^{M}h_{i}^{p}\frac{\partial}{\partial h_{i}}
 =\sum_{k=1}^{M}B_{k}^{(p)}\frac{\partial}{\partial\sigma_{k}};
 \qquad B_{k}^{(p)}=\sum_{i=1}^{M}h_{i}^{p}
 \frac{\partial\sigma_{k}}{\partial h_{i}}.
\label{eqn:form30}
\end{align}
The usefulness of the formula (\ref{eqn:form23}) is understood
if we note that the following recursion relation can be derived
from it:
\begin{align}
B_{k}^{(p+1)}&=\sum_{i=1}^{M}h_{i}^{p}\left(\sigma_{k}
 -\frac{\partial\sigma_{k+1}}{\partial h_{i}}\right)\notag\\
&=B_{-}^{(p)}\sigma_{k}-B_{k+1}^{(p)}.
\label{eqn:form31}
\end{align}
The above relation enable us to calculate all the $B_{k}^{(p)}$
needed, once one of them is known. The calculation of $B_{k}^{(1)}$
will be the easiest. Let us define a generating function
$B^{(1)}(t)$ as,
\begin{align}
B^{(1)}(t)=\sum_{k=0}^{M}B_{k}^{(1)}t^{k}.
\label{eqn:form32}
\end{align}
From the formulae (\ref{eqns:formuG}), we have another expression
for $B^{(1)}(t)$:
\begin{align}
B^{(1)}(t)&=\sum_{i=1}^{M}h_{i}\frac{\partial}{\partial h_{i}}G(t)
 =\sum_{n=1}^{\infty}(-)^{n+1}t^{n}s_{n}G(t)\notag\\
&=t\frac{\partial}{\partial t}G(t)=\sum_{k=0}^{M}k\sigma_{k}\, t^{k}.
\label{eqn:form33}
\end{align}
By comparing Eq.~(\ref{eqn:form32}) with the last expression in
Eq.~(\ref{eqn:form33}), we obtain,
\begin{align}
B_{k}^{(1)}=k\sigma_{k},
\label{eqn:form34}
\end{align}
Applying the recursion relation (\ref{eqn:form31}) to
Eq.~(\ref{eqn:form34}) successively, we finally get the formulae
for $B_{k}^{(p)}$:
\begin{subequations}
\label{eqns:formuB}
\begin{align}
B_{k}^{(0)}&=(M-k+1)\sigma_{k-1},\\
B_{k}^{(1)}&=k\sigma_{k},\\
B_{k}^{(2)}&=\sigma_{1}\sigma_{k}-(k+1)\sigma_{k+1},\\
B_{k}^{(3)}&=(\sigma_{1}^{2}-2\sigma_{2})\sigma_{k}
 -\sigma_{1}\sigma_{k+1}+(k+2)\sigma_{k+2}.
\end{align}
\end{subequations}

\subsection{\label{ssec:C-p}Calculation of $C_{-}^{(p)}(\sigma)$}

The quantity $C_{-}^{(p)}$ is defined by,
\begin{align}
\sum_{i\neq j}^{M}\frac{h_{i}^{p}}{h_{i}-h_{j}}=C_{-}^{(p)}(\sigma).
\label{eqn:for4-0}
\end{align}
A modification of Eq.~(\ref{eqn:for4-0}) reads,
\begin{align}
C_{-}^{(p)}&=\frac{1}{2}\sum_{i\neq j}^{M}
 \frac{h_{i}^{p}-h_{j}^{p}}{h_{i}-h_{j}}=\frac{1}{2}
 \sum_{i\neq j}^{M}\sum_{l=0}^{p-1}h_{i}^{l}h_{j}^{p-1-l}\notag\\
&=\frac{1}{2}\left(\sum_{l=0}^{p-1}s_{l}s_{p-1-l}-p\, s_{p-1}\right).
\label{eqn:for4-1}
\end{align}
With the aid of the formulae (\ref{eqns:formB-}) we easily obtain,
for instance,
\begin{subequations}
\label{eqns:formC-}
\begin{align}
C_{-}^{(0)}&=0,\\
C_{-}^{(1)}&=\frac{M}{2}(M-1),\\
C_{-}^{(2)}&=(M-1)\sigma_{1},\\
C_{-}^{(3)}&=(M-1)\sigma_{1}^{2}-(2M-3)\sigma_{2}.
\end{align}
\end{subequations}

\subsection{\label{ssec:Ckp}Calculation of $C_{k}^{(p)}(\sigma)$}

The quantity $C_{k}^{(p)}$ is defined by,
\begin{align}
\sum_{i\neq j}^{M}\frac{h_{i}^{p}}{h_{i}-h_{j}}
 \frac{\partial}{\partial h_{i}}=\sum_{k=1}^{M}
 C_{k}^{(p)}\frac{\partial}{\partial\sigma_{k}};
 \qquad C_{k}^{(p)}=\sum_{i\neq j}^{M}\frac{h_{i}^{p}}{h_{i}-h_{j}}
 \frac{\partial\sigma_{k}}{\partial h_{i}}.
\label{eqn:form40}
\end{align}
In order to calculate $C_{k}^{(p)}$, we define a generating function
$C^{(p)}(t)$ as,
\begin{align}
C^{(p)}(t)=\sum_{k=0}^{M}C_{k}^{(p)}t^{k}.
\label{eqn:form41}
\end{align}
From the formula (\ref{eqn:formG1}) we have,
\begin{align}
2C^{(p)}(t)&=2\sum_{k=0}^{M}C_{k}^{(p)}t^{k}=\sum_{i\neq j}^{M}
 \frac{1}{h_{i}-h_{j}}\biggl(h_{i}^{p}\frac{\del}{\del h_{i}}
 -h_{j}^{p}\frac{\del}{\del h_{j}}\biggr)G(t)\notag\\
&=\sum_{i\neq j}^{M}\sum_{m=1}^{\infty}(-)^{m+1}t^{m}\sum_{l=0}^{p+m-2}
 h_{i}^{l}h_{j}^{p+m-2-l}G(t).
\label{eqn:form42}
\end{align}
On the other hand, we obtain for $q\ge 0$ from the formula
(\ref{eqn:formG3}),
\begin{align}
t^{q}\frac{\partial^{2}}{\partial t^{2}}G(t)
=&\, G(t)\sum_{m=q}^{\infty}\sum_{l=1}^{m-q+1}(-)^{m-q+4}t^{m}
 s_{l}s_{m-q+2-l}-G(t)\sum_{m=1}^{\infty}(-)^{m+1}t^{m+q-1}
 m\, s_{m+1}\notag\\
=&\, G(t)\sum_{m=q}^{\infty}(-)^{m-q+2}t^{m}\sum_{l=0}^{m-q+2}
 s_{l}s_{m-q+2-l}\notag\\
&\, -G(t)\sum_{m=1}^{\infty}(-)^{m+1}t^{m+q-1}(2M+m)s_{m+1}.
\label{eqn:form43}
\end{align}
The coefficient of $G(t)$ in the second line of
Eq.~(\ref{eqn:form43}) is further deformed as,
\begin{align}
\sum_{m=q}^{\infty}(-)^{m-q+2}t^{m}\sum_{l=0}^{m-q+2}s_{l}s_{m-q+2-l}
=&\,\sum_{i\neq j}^{M}\sum_{m=1}^{\infty}(-)^{m-q+2}t^{m}
 \sum_{l=0}^{m-q+2}h_{i}^{l}h_{j}^{m-q+2-l}\notag\\
&\, +\sum_{m=1}^{\infty}(-)^{m+1}t^{m+q-1}(m+2)s_{m+1}+f_{q}^{C}(t),
\label{eqn:Fsubq}
\end{align}
where $f_{q}^{C}(t)$ is given by,
\begin{align}
f_{q}^{C}(t)=
\begin{cases}
(2M-3)s_{2}+s_{1}^{2}, & q=0,\\
\displaystyle{\sum_{m=1}^{q-1}}(-)^{m-q+2}t^{m}
 \displaystyle{\sum_{l=0}^{m-q+2}}(s_{m-q+2}-s_{l}s_{m-q+2-l}), & q\ge 1.
\end{cases}
\label{eqn:fCsbq}
\end{align}
Substituting Eq.~(\ref{eqn:Fsubq}) for Eq.~(\ref{eqn:form43})
and applying the formula (\ref{eqn:formG2}), we obtain,
\begin{align}
t^{q}\frac{\partial^{2}}{\partial t^{2}}G(t)
=&\,\sum_{i\neq j}^{M}\sum_{m=1}^{\infty}(-)^{m-q+2}t^{m}
 \sum_{l=0}^{m-q+2}h_{i}^{l}h_{j}^{m-q+2-l}G(t)\notag\\
&\, -2(M-1)t^{q-1}\left(-\frac{\partial}{\partial t}
 +s_{1}\right)G(t)+f_{q}^{C}(t)G(t).
\label{eqn:form44}
\end{align}
Combining Eqs.~(\ref{eqn:form41}) and (\ref{eqn:form44}), we get
another expression for $C^{(p)}(t)$ $(0\le p\le 4)$:
\begin{align}
2C^{(p)}(t)&=(-)^{p+1}\left[t^{4-p}\frac{\del^{2}}{\del t^{2}}-2(M-1)
 t^{3-p}\biggl(\frac{\del}{\del t}-\sigma_{1}\biggr)-f_{4-p}^{C}(t)
 \right]G(t)\notag\\
&=(-)^{p+1}\sum_{k=0}^{M}\bigl[k(k-2M+1)t^{2-p}+2(M-1)\sigma_{1}t^{3-p}
 -f_{4-p}^{C}(t)\bigr]\sigma_{k}t^{k}.
\label{eqn:form45}
\end{align}
Comparison of Eq.~(\ref{eqn:form41}) with Eq.~(\ref{eqn:form45})
enable us to calculate $C_{k}^{(p)}$ for $0\le p\le 4$.
The recursion relation derived from Eq.~(\ref{eqn:form23}) in
this case is,
\begin{align}
C_{k}^{(p+1)}=\sigma_{k}C_{-}^{(p)}-C_{k+1}^{(p)}.
\label{eqn:form46}
\end{align}
The calculation of $C_{k}^{(2)}$ may be the easiest. From
Eqs.~(\ref{eqn:form45}) and (\ref{eqn:form46}), we obtain
the formulae for $C_{k}^{(p)}$:
\begin{subequations}
\label{eqns:formuC}
\begin{align}
2C_{k}^{(0)}=&\,-(k-M-1)(k-M-2)\sigma_{k-2},\\
2C_{k}^{(1)}=&\,(k-M)(k-M-1)\sigma_{k-1},\\
2C_{k}^{(2)}=&\,-k(k-2M+1)\sigma_{k},\\
2C_{k}^{(3)}=&\,2(M-1)\sigma_{1}\sigma_{k}+(k+1)(k-2M+2)\sigma_{k+1},\\
2C_{k}^{(4)}=&\,2(M-1)\sigma_{1}^{2}\sigma_{k}-2(2M-3)\sigma_{2}\sigma_{k}
 -2(M-1)\sigma_{1}\sigma_{k+1}\notag\\
&\,-(k+2)(k-2M+3)\sigma_{k+2}.
\end{align}
\end{subequations}

\subsection{\label{ssec:Aklp}Calculation of $A_{k,l}^{(p)}(\sigma)$}

The definition of the quantity $A_{k,l}^{(p)}$ is given by,
\begin{align}
\sum_{i=1}^{M}h_{i}^{p}\frac{\partial^{2}}{\partial h_{i}^{2}}
 =\sum_{k,l=1}^{M}A_{k,l}^{(p)}\frac{\partial^{2}}%
{\partial\sigma_{k}\partial\sigma_{l}};\qquad A_{k,l}^{(p)}=
 \sum_{i=1}^{M}h_{i}^{p}\frac{\partial\sigma_{k}}{\partial h_{i}}
 \frac{\partial\sigma_{l}}{\partial h_{i}}.
\label{eqn:form50}
\end{align}
In order to calculate $A_{k,l}^{(p)}$, we introduce a generating
function $A^{(p)}(t,u)$ defined by,
\begin{align}
A^{(p)}(t,u)=\sum_{k,l=0}^{M}A_{k,l}^{(p)}t^{k}u^{l}.
\label{eqn:form51}
\end{align}
Then, it immediately reads,
\begin{align}
A^{(p)}(t,u)
&=\sum_{i=1}^{M}h_{i}^{p}\frac{\partial G(t)}{\partial h_{i}}
 \frac{\partial G(u)}{\partial h_{i}}\notag\\
&=\sum_{m',n'=1}^{\infty}(-)^{m'+n'+2}s_{m'+n'+p-2}
 t^{m'}u^{n'}G(t)G(u).
\label{eqn:form52}
\end{align}
If we make the substitution $m=m'+n'$ and $n=m'-n'$, the expression
of $A^{(p)}(t,u)$ is rewritten as,
\begin{align}
A^{(p)}(t,u)&=\sum_{m=2}^{\infty}(-)^{m}s_{m+p-2}(tu)^{m/2}
 \sum_{n=2-m}^{m-2}\left(\frac{t}{u}\right)^{n/2}G(t)G(u)\notag\\
&=\frac{tu}{t-u}\sum_{m=1}^{\infty}(-)^{m+1}s_{m+p-1}
 (t^{m}-u^{m})G(t)G(u).
\label{eqn:form53}
\end{align}
On the other hand, utilizing the formula (\ref{eqn:formG2}) we have
for $p\ge 0$,
\begin{align}
t^{2-p}\frac{\partial}{\partial t}G(t)&=\sum_{m=1}^{\infty}
 (-)^{m+1}t^{m-p+1}s_{m}G(t)\notag\\
&=(-)^{p+1}\sum_{m=1}^{\infty}(-)^{m+1}t^{m}s_{m+p-1}G(t)
 +f_{p}^{A}(t)G(t),
\label{eqn:form54}
\end{align}
where $f_{p}^{A}(t)$ is given by,
\begin{align}
f_{p}^{A}(t)=
\begin{cases}
Mt, & p=0,\\
\displaystyle{\sum_{m=1}^{p-1}}(-)^{m+1}t^{m-p+1}s_{m}, & p\ge 1.
\end{cases}
\label{eqn:fAsbq}
\end{align}
From Eqs.~(\ref{eqn:form53}) and (\ref{eqn:form54}), we obtain
another expression for $A^{(p)}(t,u)$:
\begin{align}
A^{(p)}(t,u)
&=(-)^{p+1}\frac{tu}{t-u}\left[t^{2-p}\frac{\partial}{\partial t}
 -u^{2-p}\frac{\partial}{\partial u}-f_{p}^{A}(t)+f_{p}^{A}(u)
 \right]G(t)G(u)\notag\\
&=\frac{(-)^{p+1}}{t-u}\sum_{k,l=0}^{M}\left[k\, t^{1-p} -l\, u^{1-p}
 -f_{p}^{A}(t)+f_{p}^{A}(u)\right]\sigma_{k}\sigma_{l}\, t^{k+1}u^{l+1}.
\label{eqn:form55}
\end{align}
Comparison of Eq.~(\ref{eqn:form51}) with Eq.~(\ref{eqn:form55})
enable us to calculate $A_{k,l}^{(p)}$. The recursion relation
derived from the formula (\ref{eqn:form23}) in this case is,
\begin{align}
A_{k,l}^{(p+1)}=B_{k}^{(p)}\sigma_{l}-A_{k,l+1}^{(p)}
 =\sigma_{k}B_{l}^{(p)}-A_{k+1,l}^{(p)}.
\label{eqn:form56}
\end{align}
The calculation of $A_{k,l}^{(1)}$ may be the easiest.
From Eqs.~(\ref{eqn:form55}) and (\ref{eqn:form56}) we obtain,
\begin{subequations}
\label{eqns:formuA}
\begin{align}
A_{k,l}^{(0)}=&\,(M-l+1)\sigma_{k-1}\sigma_{l-1}+\sum_{m=0}^{k-2}
 (k-l-2m-2)\sigma_{k-m-2}\sigma_{l+m},\\
A_{k,l}^{(1)}=&\,-\sum_{m=0}^{k-1}(k-l-2m-1)
 \sigma_{k-m-1}\sigma_{l+m},\\
A_{k,l}^{(2)}=&\,k\sigma_{k}\sigma_{l}+\sum_{m=1}^{k}(k-l-2m)
 \sigma_{k-m}\sigma_{l+m},\\
A_{k,l}^{(3)}=&\,\sigma_{1}\sigma_{k}\sigma_{l}-\sum_{m=0}^{1}
(k-m+1)\sigma_{k-m+1}\sigma_{l+m}-\sum_{m=2}^{k+1}(k-l-2m+1)
 \sigma_{k-m+1}\sigma_{l+m},\\
A_{k,l}^{(4)}
=&\,\left(\sigma_{1}^{2}-2\sigma_{2}\right)\sigma_{k}\sigma_{l}
 -\sigma_{1}\left(\sigma_{k+1}\sigma_{l}+\sigma_{k}\sigma_{l+1}
 \right)\notag\\
&\, +\sum_{m=0}^{2}(k-m+2)\sigma_{k-m+2}\sigma_{l+m}
 +\sum_{m=3}^{k+2}(k-l-2m+2)\sigma_{k-m+2}\sigma_{l+m}.
\end{align}
\end{subequations}


\end{document}